\documentclass{ws-ijmpe}
\usepackage[super,compress]{cite}
\begin{document}

\markboth{P.~Van~Isacker}{Neutron--Proton Pairs in Nuclei}

\catchline{}{}{}{}{}

\title{NEUTRON--PROTON PAIRS IN NUCLEI}

\author{\footnotesize P.~VAN~ISACKER}

\address{Grand Acc\'el\'erateur National d'Ions Lourds, CEA/DSM--CNRS/IN2P3\\
B.P.~55027, F-14076 Caen Cedex 5, France\\
isacker@ganil.fr}

\maketitle

\begin{history}
\received{Day Month Year}
\revised{Day Month Year}
\end{history}

\begin{abstract}
A review is given of attempts to describe nuclear properties
in terms of neutron--proton pairs
that are subsequently replaced by bosons.
Some of the standard approaches with low-spin pairs are recalled
but the emphasis is on a recently proposed framework
with  pairs of neutrons and protons with aligned angular momentum.
The analysis is carried out for general $j$
and applied to $N=Z$ nuclei in the $1f_{7/2}$ and $1g_{9/2}$ shells.
\end{abstract}


\ccode{PACS numbers: 03.65.Fd,21.60.Cs, 21.60.Ev}


\section{Introduction}
\label{s_intro}
In dealing with complex systems with many elementary components,
one of the major goals of physics
is to seek simplifications by adopting a description in terms of composite structures.
An obvious example of this approach is found in nuclear physics,
when the elementary constituents of the nucleus, quarks,
are lumped into nucleons---an approximation
adequate for the description of most nuclear phenomena at low energy.
Still, the nuclear many-body problem in terms of nucleons instead of quarks
is fiendishly difficult to solve for all but the lightest of nuclei,
and further simplifying assumptions are required for the majority of them.
One possibility is to lump the nucleons into pairs
and attempt a description of nuclear phenomena in terms of those.

While such nucleon-pair models can be simple and attractive in principle,
their success obviously depends on the type of pairs considered.
This choice should be guided by nature
of the interaction between the nucleons.
One of the defining features of the nuclear force
is that it is strongly attractive between nucleons
that are paired to angular momentum $J=0$.
Models where the pairing component of the interaction is prominent
therefore have played
an important part in the development of our understanding of nuclear structure~\cite{Brink05}.
While pairs consisting of {\em identical} nucleons
({\it i.e.}, neutron--neutron or proton--proton)
are by now an accepted feature of nuclei,
much debate still exists concerning the role of neutron--proton pairs.
One component of neutron--proton pairing is of isovector character,
and arguments of isospin symmetry require
that it should be considered on the same footing
as its neutron--neutron or proton--proton equivalent.
A neutron and a proton can also interact
via an isoscalar component of the nuclear force
and the debate is whether pairing of this type
leads to enhanced collectivity and correlated states.
This question is still unanswered
after several decades of
research~\cite{Frobrich71,Engel96,Macchiavelli00,Vogel00,Chasman02,Isacker05,Bertsch10}.
A recent review of possible signatures
of isoscalar neutron--proton pairing is given by Macchiavelli~\cite{Macchiavelli12}.

This paper certainly does {\em not} give
a comprehensive and exhaustive review of the role of neutron--proton pairs in nuclei.
Rather, it zooms in on a particular approach
which replaces pairs of nucleons by bosons---approximation
known under the name of `interacting boson model'---
and within this class of models attention is paid to those
that adopt bosons that stem from neutron--proton pairs.
The standard boson models of this kind
are briefly described in Sect.~\ref{s_ibm}
but the emphasis is on a recently proposed framework
with bosons that correspond to neutron--proton pairs
with aligned, high angular momentum.
The motivation for and the historical context of this new approach
are outlined in Sect.~\ref{s_nppairs}.
The main purpose of the present review
is to argue that the proper framework to develop this approach
is by applying boson mapping techniques to the nucleon-pair shell model.
Technical aspects are reviewed in Sects.~\ref{s_npsm} and~\ref{s_mapping}
while Sect.~\ref{s_approx} gives a non-technical summary
of the various approximations that enter a description
in terms of aligned neutron--proton pairs or bosons.
Applications to the $1f_{7/2}$ and $1g_{9/2}$ shells
are discussed in Sect.~\ref{s_appli}.
In fact, most of the results shown in that section are new
and in this sense the present paper is not a review of published research.
Nevertheless, it is the opinion of the author
that they clarify the issue of the role in nuclei
of neutron--proton pairs, aligned or otherwise.
Finally, conclusions are drawn in Sect.~\ref{s_conc}. 

\section{Standard boson models with neutron--proton pairs: IBM-3 and IBM-4}
\label{s_ibm}
The interacting boson model (IBM) of Arima and Iachello~\cite{Arima75}
starts from the premise that low-lying collective excitations
can be described in terms of nucleon pairs
(with angular momentum $0$ and $2$,
in the most elementary version of the model)
and that these pairs can be approximated as ($s$ and $d$) bosons.
If neutrons and protons occupy different valence shells,
it is natural to consider neutron--neutron and proton--proton pairs only,
and to include the neutron--proton interaction
as a force between the two types of pairs.
This then leads to a version of the IBM with two kinds of bosons~\cite{Arima77},
of neutron and of proton type,
the so-called \mbox{IBM-2}.
If neutrons and protons occupy the same valence shell,
this approach is no longer valid
since there is no reason not to include a neutron--proton pair
with isospin $T=1$.
The version of the IBM that also contains the $T=1$ neutron--proton boson,
proposed by Elliott and White~\cite{Elliott80}, is called \mbox{IBM-3}.
Because the \mbox{IBM-3} includes the complete $T=1$ triplet,
it can be made isospin invariant,
enabling the construction of states
with good total angular momentum $J$ {\em and} good total isospin $T$
and leading therefore to a more direct comparison with the shell model
(see, {\it e.g.}, Ref.~\refcite{Thompson87}).

The bosons of the \mbox{IBM-3} all have isospin $T=1$
and, in principle, other bosons can be introduced,
in particular those that correspond to $T=0$ neutron--proton pairs.
This further extension
(proposed by Elliott and Evans~\cite{Elliott81} and referred to as \mbox{IBM-4})
is the most elaborate version of the standard IBM.
The bosons are assigned an orbital angular momentum $L$, a spin $S$ and an isospin $T$,
and in \mbox{IBM-4} the choice $L=0$ and $L=2$ is retained
with either $(S,T)=(0,1)$ or $(S,T)=(1,0)$.
The total angular momentum $J$ of the bosons
is obtained by coupling $L$ and $S$,
leading to an ensemble of bosons with $(J,T)=(0,1)$, $(2,1)$, $(1,0)^2$, $(2,0)$ and $(3,0)$.

\begin{figure}
\includegraphics[height=5cm]{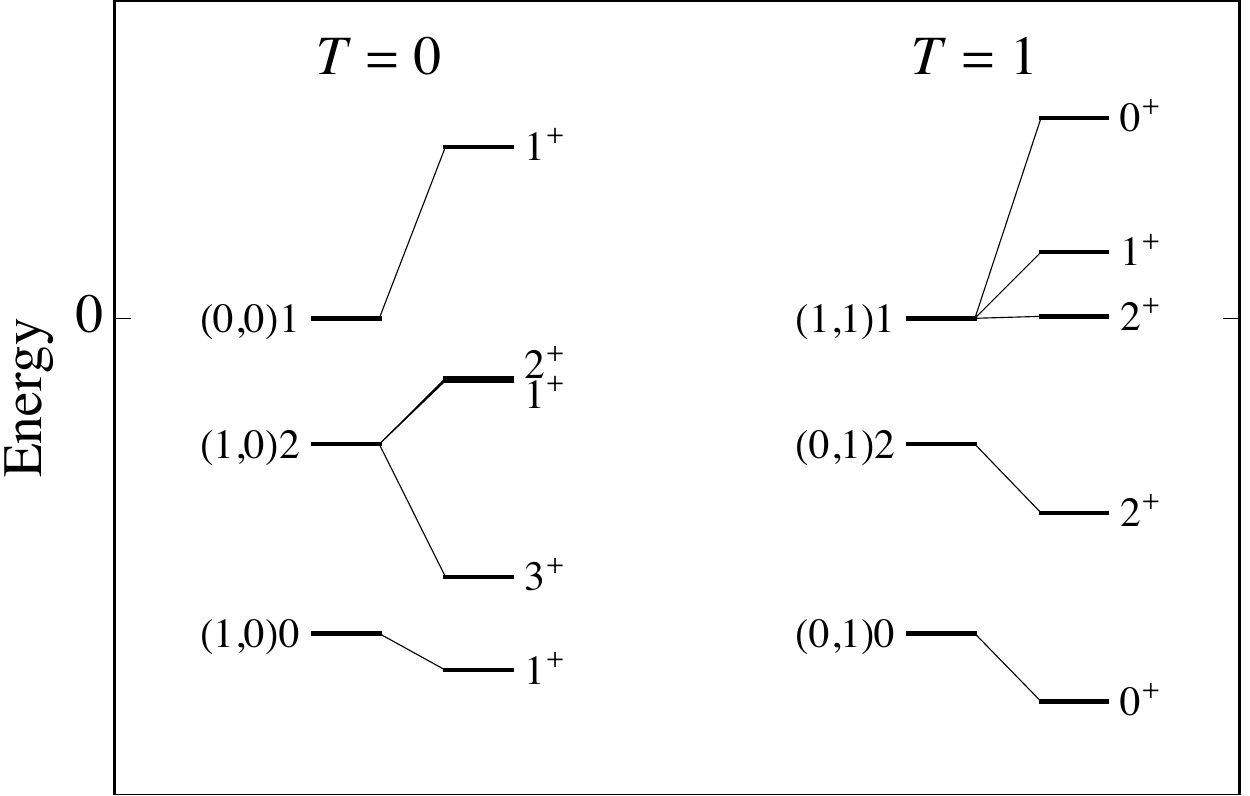}
\caption{
\label{f_npspec}
The energy spectrum of a neutron and a proton in a $p$ shell,
interacting through an attractive delta force.
States with isospin $T=0$ ($T=1$)
are shown on the left (right).
Levels are labelled on the left by $(S,T)L$,
the spin $S$, the isospin $T$ and the orbital angular momentum $L$,
and on the right by the angular momentum and parity $J^\pi$.
All these quantum numbers are conserved
in the absence of a spin--orbit splitting (left)
while only the $J^\pi$ symmetry remains for a non-zero spin--orbit splitting (right).}
\end{figure}
The justification of this particular choice of bosons is based on the shell model.
Consider as an example a neutron and a proton in a $p$ shell.
The effective force between the two nucleons is of a short-range nature
and can, within a reasonable approximation, be represented
as an attractive delta interaction,
$\hat V(\bar r_1-\bar r_2)=-g\delta(\bar r_1-\bar r_2)$ with $g>0$.
Under the assumption of zero spin--orbit splitting
({\it i.e.}, degenerate $p_{1/2}$ and $p_{3/2}$ shells),
the energy spectrum can be worked out on the basis of simple symmetry arguments
(see Fig.~\ref{f_npspec}).
Since the interaction is spin and isospin independent,
the $LS$ coupling scheme applies and all states
can be assigned an orbital angular momentum $L$, a spin $S$ and an isospin $T$.
Furthermore, because of overall anti-symmetry,
all states are characterized by
either spatial symmetry ($L=0$ or 2) and spin--isospin anti-symmetry [$(S,T)=(0,1)$ or $(1,0)$],
or spatial anti-symmetry ($L=1$) and spin--isospin symmetry [$(S,T)=(0,0)$ or $(1,1)$].
The former states are lowered in energy by the attractive delta force
($L=0$ more so than $L=2$)
while the interaction energy in the latter states is exactly zero
because of their spatial anti-symmetry.
The states lowered in energy are precisely
those that correspond to the bosons in \mbox{IBM-4}.
For a realistic choice of spin--orbit splitting,
the many degeneracies are lifted,
lowering the higher-$J$ levels in energy (see Fig.~\ref{f_npspec}).
The choice of bosons in \mbox{IBM-4} allows a classification
where states carry the quantum numbers
of total orbital angular momentum $L$,
total spin $S$, total angular momentum $J$ and total isospin $T$,
in addition to the SU(4) labels $(\lambda,\mu,\nu)$ of Wigner's supermultiplet scheme~\cite{Wigner37},
in close analogy with the corresponding shell-model labels.

\begin{figure}
\includegraphics[height=5cm]{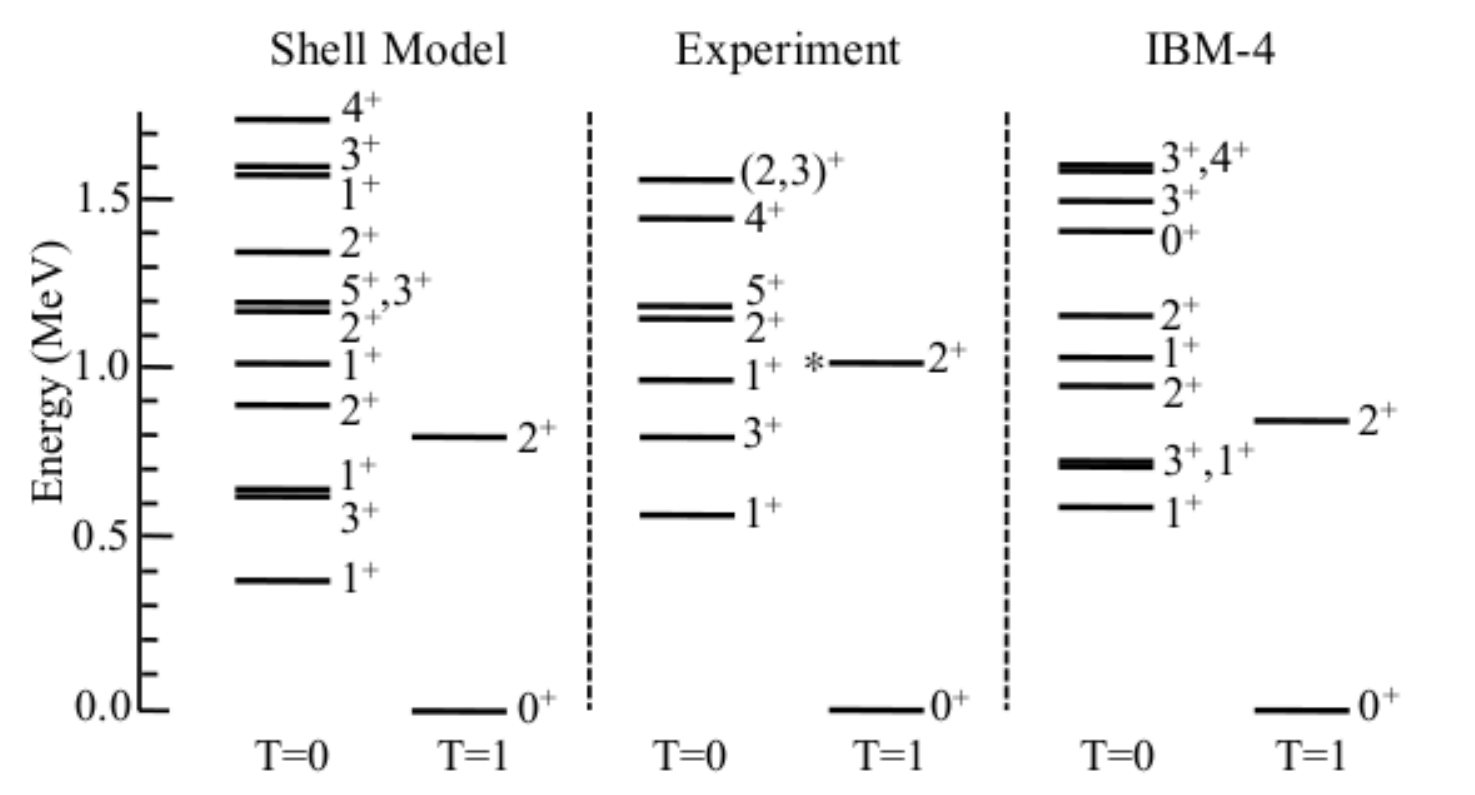}
\caption{
\label{f_ga62}
The spectrum of states in $^{62}$Ga with isospin $T=0$ and $T=1$.
Levels are labelled by their angular momentum and parity $J^\pi$.
The different columns contain
the results of a large-scale shell-model calculation~\cite{Vincent98},
of a mapped \mbox{IBM-4} calculation~\cite{Juillet01}
and the experimental levels~\cite{David13un}.
Figure taken from David {\it et al.}~\cite{David13un}.}
\end{figure}
These qualitative arguments in favour of the \mbox{IBM-4}
have been corroborated by quantitative, microscopic studies
in even--even~\cite{Halse84}
and odd--odd~\cite{Halse85} $sd$-shell nuclei.
In heavier nuclei the situation is more complex.
The effect of the spin--orbit force is such
that the $LS$-coupling scheme no longer applies,
resulting in the breaking of the $L$ and $S$ quantum numbers,
in contrast to the total angular momentum $J$
which is of course exactly conserved because of rotational invariance
and the total isospin $T$ which conserved to a good approximation.
Nevertheless, the $L$ and $S$ quantum numbers of the shell model
can be replaced by their `pseudo' equivalents,
along the original ideas of Hecht and Adler~\cite{Hecht69},
and Arima {\it et al.}~\cite{Arima69}.
This might be possible in specific regions of the nuclear chart~\cite{Strottman72}
and is borne out by shell-model calculations with realistic interactions
in nuclei beyond $^{56}$Ni~\cite{Isacker99}.
The existence of these approximate symmetries in the shell model
allows a mapping onto IBM-4.
A typical example is provided by the $N=Z$ nucleus $^{62}$Ga.
The spectroscopy predicted in the shell model,
with a space consisting of the orbits $2p_{3/2}$, $1f_{5/2}$, $2p_{1/2}$ and $1g_{9/2}$,
is very complex with intertwined states of isospin $T=0$ and $T=1$~\cite{Vincent98}.
Most levels are of low spin and those are well reproduced
in the mapped \mbox{IBM-4} calculation~\cite{Juillet01}.
It is also found, however, that $T=0$ levels with higher spin ($5^+$, $6^+$ and $7^+$)
are at significantly higher energies in the \mbox{IBM-4}
or even absent from it.
This result is not surprising
since the standard \mbox{IBM-4} choice
consists of bosons with rather low spin (up to $J^\pi=3^+$).
In a recent experiment,
David {\it et al.}~\cite{David13un} have observed
a number of additional levels of low spin,
presumably with isospin $T=0$,
as predicted by the shell model and the \mbox{IBM-4} (see Fig.~\ref{f_ga62}).
It remains nevertheless true that states with higher spin
require an approach which is different from the standard \mbox{IBM-4}.

\section{Aligned neutron--proton pairs}
\label{s_nppairs}
In a recent paper, Cederwall {\it et al.}~\cite{Cederwall11}
propose an alternative description of $N\sim Z$ nuclei
in terms of neutron--proton pairs with {\em aligned} spin,
henceforth referred to as $B$ pairs.
The proposal concerns massive $N\sim Z$ nuclei,
such as $^{92}$Pd, approaching $^{100}$Sn,
with valence nucleons dominantly in the $1g_{9/2}$ shell.
The claim is made (see also Refs.~\refcite{Qi11} and \refcite{Xu12})
that low-lying yrast states in $^{92}$Pd and neighbouring nuclei
are mainly built out of aligned neutron--proton isoscalar (with isospin $T=0$) pairs
with angular momentum $J=9$.
For the purpose of constructing a boson model,
the aligned-pair scheme is particularly attractive
since it involves a single neutron--proton pair;
if valid, the many bosons of \mbox{IBM-4}
can be replaced by a single one.

Related ideas have been explored in the past.
One is the stretch scheme of Danos and Gillet~\cite{Danos66,Danos67}
which applies to even--even $N=Z$ nuclei.
It assumes that half of the neutrons align with half of the protons
to form a state of maximum angular momentum.
Similarly, the other half of the nucleons aligns to a state
with the same angular momentum.
The total angular momentum of the system
is generated by the coupling of these two fixed stretched configurations.
For four nucleons the stretch scheme is exactly equivalent
to the description in terms of aligned pairs
as proposed by Blomqvist and co-workers~\cite{Cederwall11}.
For eight, twelve,\dots nucleons, however, the stretch scheme is different
since any angular momentum is uniquely defined
in terms of the two stretch configurations
while it generally can be written in several ways in terms of $B$ pairs.
As a result, the stretch scheme has less flexibility
to provide an adequate approximation of a realistic shell-model wave function.
An explicit relation between both approximations
is established in Sect.~\ref{s_appli}.

Blomqvist's aligned $B$ pairs are in fact identical to the `$q$ pairs'
introduced in the 1980s by Daley.
A $q$-pair analysis of the $1f_{7/2}$ shell with a schematic delta interaction
exists as a Daresbury preprint~\cite{Daley87}
but, unfortunately, not as a published paper.
The study of Daley concentrates on the even--even nuclei $^{44}$Ti and $^{48}$Cr,
and only in the former nucleus
does he find results similar to the ones shown below.
No analysis of odd--odd nuclei is presented.

A crucial issue in any model that represents a fermionic system
in terms of pairs (or, more generally, clusters) of fermions
is the representation of exchange effects resulting from the Pauli principle
as interactions between these clusters.
In the stretch scheme of Danos and Gillet~\cite{Danos66,Danos67}
anti-symmetry between the two stretched configurations is neglected
while it is not clear from Daley's paper~\cite{Daley87}
to what extent the interactions between his $q$ bosons include Pauli effects.
On the other hand, anti-symmetry is fully taken into account
in the multi-step shell-model approach of Qi {\it et al.}~\cite{Qi11},
at the expense of major computational complexities
which hinder an easy, intuitive interpretation of the results.
One of the aims of this review is to analyze
results of shell-model calculations in terms of $B$ pairs
with the nucleon-pair shell model.
Although numerically challenging,
in particular for $B$ pairs in view of their high angular momentum,
this approach provides a conceptually simple way
to treat Pauli exchange effects between the pairs
and subsequently represent those as interactions between bosons.
The technical aspects of this approach are reviewed in the next two sections.

\section{Nucleon-pair shell model}
\label{s_npsm}
The natural framework to test Blomqvist's truncation idea
is provided by the nucleon-pair shell model (NPSM)~\cite{Chen93,Chen97}.
In the NPSM a basis is constructed from nucleon pairs.
These can be collective superpositions of two-particle states
or they may be identified with pure two-particle states themselves.
The many applications of this formalism are reviewed in Ref.~\refcite{Zhao13un}.
The extension of the NPSM that includes isospin~\cite{Fu13a}
is of particular relevance here.

In the language and notation of the NPSM,
Blomqvist's idea can be summarized with the statement
that the full $T=0$ shell-model space can be reduced to one
constructed out of aligned neutron--proton pairs
of which the basis states are written as
\begin{equation}
|B^nL_2\dots L_n\rangle\equiv
\left(\cdots\left(\left(B^\dag\times B^\dag\right)^{(L_2)}\times B^\dag\right)^{(L_3)}\times\cdots\times
B^\dag\right)^{(L_n)}|{\rm o}\rangle,
\label{e_fnobas}
\end{equation}
with $|{\rm o}\rangle$ the vacuum state.
Pairs with angular momentum $J$ and projection $M_J$,
and with isospin $T$ and projection $M_T$
are denoted by
\begin{equation}
P^\dag_{JM_JTM_T}\equiv
(a^\dag_{jt}\times a^\dag_{jt})^{(JT)}_{M_JM_T},
\label{e_fpair}
\end{equation}
where $a^\dag_{jm_jtm_t}$ creates a nucleon
with angular momentum $j$ and projection $m_j$,
and isospin $t=\frac12$ and projection $m_t$.
The short-hand notation $B^\dag_{M_J}$ ($B$ for Blomqvist) is used in Eq.~(\ref{e_fnobas})
for a creation operator of a neutron--proton pair
with angular momentum $J=2j$ and isospin $T=0$.
The $2n$-particle state~(\ref{e_fnobas}) is characterized
by the intermediate angular momenta $L_2,\dots,L_n$,
where $L_n$ is the final and total angular momentum of the state.
In the basis~(\ref{e_fnobas}) all pairs have $T=0$
and the coupling in isospin need not be considered.

The basis~(\ref{e_fnobas}) is non-orthogonal
and possibly overcomplete.
Any calculation in this basis must therefore start
from the diagonalization of the overlap matrix
\begin{equation}
\langle B^n_i|B^n_{i'}\rangle\equiv
\langle B^nL_2\dots L_{n-1}L_n|B^nL'_2\dots L'_{n-1}L_n\rangle,
\label{e_fover}
\end{equation}
where in bra and ket of the matrix element
all possible intermediate angular momenta $L_2,\dots,L_{n-1}$ must be considered,
leading to a series of basis states denoted as $|B^n_i\rangle$
where $i$ is a short-hand notation for the set $\{L_2,\dots,L_{n-1}\}$.
The computation of the matrix elements~(\ref{e_fover}) is complicated
but possible with the recurrence relation devised by Chen~\cite{Chen97}.
Vanishing eigenvalues of the overlap matrix $\langle B^n_i|B^n_{i'}\rangle$
indicate the overcompleteness of the pair basis.
If a selection of $\omega$ pair-basis states is made
for which all eigenvalues of the overlap matrix are non-zero,
the following linear combinations can be constructed:
\begin{equation}
|\bar B^n_k\rangle=
\sqrt{\frac{1}{o_k}}
\sum_{i=1}^\omega c_{ki}|B^n_i\rangle,
\qquad
k=1,\dots,\omega,
\label{e_fobas}
\end{equation}
where $o_k$ is the $k^{\rm th}$ eigenvalue of the overlap matrix
and $c_{ki}$ $(i=1,\dots,\omega)$ the associated eigenvector.
The vectors $|\bar B^n_k\rangle$ $(k=1,\dots,\omega)$
are normalized, orthogonal and linearly independent,
and therefore provide a proper basis for a shell-model calculation,
albeit a truncated one.
For a given shell-model hamiltonian $\hat H^{\rm f}$,
the energy spectrum and eigenvectors can be obtained
from the diagonalization of the matrix
\begin{equation}
\langle\bar B^n_k|\hat H^{\rm f}|\bar B^n_{k'}\rangle=
\sqrt{1\over{o_ko_{k'}}}
\sum_{i,i'=1}^\omega c_{ki}c_{k'i'}
\langle B^n_i|\hat H^{\rm f}|B^n_{i'}\rangle,
\qquad
k,k'=1,\dots,\omega.
\label{e_fme}
\end{equation}

The formalism as explained so far
allows one to perform a shell-model calculation
in a truncated basis constructed from aligned $T=0$ neutron--proton pairs.
In subsequent applications we will also want
to analyze arbitrary shell-model wave functions in terms of $B$ pairs.
An analysis of this type clearly cannot be carried out in the basis~(\ref{e_fnobas})---since
the latter spans only part of the shell-model space---and
it requires a generalization to a basis in terms of arbitrary pairs.
The formalism of the NPSM with isospin, needed to this end, is detailed in Ref.~\refcite{Fu13a}
and only a few basic formulas are given here.

It is convenient to introduce the following short-hand notation for the pairs:
\begin{equation}
P^\dag_{\Gamma M_\Gamma}\equiv
P^\dag_{JM_JTM_T}\equiv
(a^\dag_\gamma\times a^\dag_\gamma)^{(\Gamma)}_{M_\Gamma}\equiv
(a^\dag_{jt}\times a^\dag_{jt})^{(JT)}_{M_JM_T},
\label{e_fpairg}
\end{equation}
where $\gamma$ stands for $jt$, $\Gamma$ for $JT$,
$m_\gamma$ for $m_jm_t$ and $M_\Gamma$ for $M_JM_T$.
An arbitrary pair state can then be written as
\begin{equation}
|\Gamma_1\dots \Gamma_n\Lambda_2\dots \Lambda_n\rangle\equiv
\left(\cdots\left(\left(P_{\Gamma_1}^\dag\times P_{\Gamma_2}^\dag\right)^{(\Lambda_2)}\times
P_{\Gamma_3}^\dag\right)^{(\Lambda_3)}\times\cdots\times
P_{\Gamma_n}^\dag\right)^{(\Lambda_n)}|{\rm o}\rangle,
\label{e_fnobasg}
\end{equation}
which can be denoted in short as
\begin{equation}
|P^n_j\rangle\equiv|\Gamma_1\dots \Gamma_n\Lambda_2\dots \Lambda_n\rangle,
\label{e_fnobasg2}
\end{equation}
where the index $j$ stands for the set $\{\Gamma_1\dots \Gamma_n\Lambda_2\dots \Lambda_{n-1}\}$,
that is, the angular momenta and isospins $\Gamma_q$ of the $n$ pairs,
and the intermediate angular momenta and isospins $\Lambda_q$.
Note that $\Lambda_1$ (not shown) equals $\Gamma_1$
and that $\Lambda_n$ is the total angular momentum and isospin,
and therefore fixed and not included in $j$.
Since Chen's algorithm~\cite{Chen97} is valid for arbitrary pairs,
the analysis now proceeds as before,
and consists of the construction of an orthonormal basis
from the diagonalization of the overlap matrix $\langle P^n_j|P^n_{j'}\rangle$,
\begin{equation}
|\bar P^n_r\rangle=
\sqrt{1\over{O_r}}
\sum_{j=1}^\Omega C_{rj}|P^n_j\rangle,
\qquad
r=1,\dots,\Omega,
\label{e_fobasg}
\end{equation}
where $O_r$ and $C_{rj}$ have the same meaning as in Eq.~(\ref{e_fobas})
but now in the full shell-model basis of dimension $\Omega$.
The diagonalization of the shell-model hamiltonian in that basis,
\begin{equation}
\langle\bar P^n_r|\hat H^{\rm f}|\bar P^n_{r'}\rangle=
\sqrt{1\over{O_rO_{r'}}}
\sum_{j,j'=1}^\Omega C_{rj}C_{r'j'}
\langle P^n_j|\hat H^{\rm f}|P^n_{j'}\rangle,
\qquad
r,r'=1,\dots,\Omega,
\label{e_fmeg}
\end{equation}
leads to the untruncated eigenspectrum of the shell model.

The $B$-pair content of an arbitrary shell-model state
can now be analyzed as follows.
First, a shell-model diagonalization is performed
in a complete basis $|\bar P^n_r\rangle$ $(r=1,\dots,\Omega)$,
leading to eigenstates
\begin{equation}
|\bar E^n_s\rangle=
\sum_{r=1}^\Omega E_{sr}|\bar P^n_r\rangle,
\qquad
s=1,\dots,\Omega.
\label{e_fanal1}
\end{equation}
The $B$-pair content of a given eigenstate $|\bar E^n_s\rangle$
is the square of its projection onto the subspace spanned by $B$-pair states
which equals
\begin{equation}
\langle\bar E^n_s|B^n\rangle^2\equiv
\sum_{k=1}^\omega|\langle\bar E^n_s|\bar B^n_k\rangle|^2,
\label{e_fana}
\end{equation}
where the overlap matrix element on the right-hand side
can expressed as
\begin{equation}
\langle\bar E^n_s|\bar B^n_k\rangle=
\sum_{r,j=1}^\Omega\sum_{i=1}^\omega
E_{sr}\sqrt{\frac{1}{O_r}}C_{rj}\sqrt{\frac{1}{o_k}}c_{ki}
\langle P^n_j|B^n_i\rangle,
\label{e_fanal2}
\end{equation}
in terms of overlap matrix elements that can be computed with Chen's algorithm~\cite{Chen97}.

A final word is needed concerning the calculation
of matrix elements of a shell-model hamiltonian between pair states
as they appear on the right-hand sides of Eqs.~(\ref{e_fme}) and~(\ref{e_fmeg}).
For the case of a single-$j$ shell,
the one-body part of $\hat H^{\rm f}$ gives rise to a constant
and can be neglected.
Its two-body part $\hat H^{\rm f}_2$
is entirely determined by the two-body matrix elements
\begin{equation}
\upsilon^{2\rm f}_\Gamma\equiv
\upsilon^{2\rm f}_{JT}\equiv
\langle j^2JT|\hat H^{\rm f}_2|j^2JT\rangle,
\label{e_sm2}
\end{equation}
which enter as follows in the expression for the pair matrix element:
\begin{eqnarray}
\lefteqn{\langle\Gamma_1\dots \Gamma_n\Lambda_2\dots \Lambda_n|
\hat H^{\rm f}_2
|\Gamma'_1\dots \Gamma'_n\Lambda'_2\dots \Lambda'_n\rangle}
\nonumber\\&=&
\delta_{\Lambda_n\Lambda'_n}
\left[\frac{4n-2\gamma-1}{2\gamma+1}
\langle\Gamma_1\dots \Gamma_n\Lambda_2\dots \Lambda_n
|\Gamma'_1\dots \Gamma'_n\Lambda'_2\dots \Lambda'_n\rangle
\sum_\Gamma(2\Gamma+1)\upsilon^{2\rm f}_\Gamma+\right.
\nonumber\\&&\left.
\sum_{\Gamma\Lambda}
\frac{2\Lambda+1}{2(2\Lambda_n+1)}
\langle\Gamma_1\dots \Gamma_n\Gamma\Lambda_2\dots \Lambda_n\Lambda
|\Gamma'_1\dots \Gamma'_n\Gamma\Lambda'_2\dots \Lambda'_n\Lambda\rangle
\upsilon^{2\rm f}_\Gamma\right],
\label{e_fme2g}
\end{eqnarray}
with $2\gamma+1=2(2j+1)$, $2\Gamma+1=(2J+1)(2T+1)$, $2\Lambda_q+1=(2L_q+1)(2T_q+1)$, and so on. 
The second sum is over all possible pairs with angular momentum and isospin $\Gamma$
which couples with $\Lambda_n$ to all possible $\Lambda$,
the total angular momentum and isospin of the $(n+1)$-pair state.

It is well known~\cite{Poves81}
that the limitation to a restricted model space ({\it e.g.}, a single-$j$ shell)
leads to an effective hamiltonian with higher-order interactions
(see Refs.~\citen{Volya09} and \citen{Isacker10} for a recent discussion
of $T=1$ three-body interactions in the $1f_{7/2}$ shell).
Equations~(\ref{e_fme}) and~(\ref{e_fmeg}) are generally valid,
irrespective of the order of the interaction in $\hat H^{\rm f}$.
Equation~(\ref{e_fme2g}), on the other hand,
is specific to a two-body interaction
but it can be readily generalized to higher orders.
The corresponding expression for a three-body interaction, for example,
involves the same overlap matrix elements as those in Eq.~(\ref{e_fme2g})
with in addition overlaps between states of $(n+1)$ pairs plus one particle.
These can be computed with the NPSM algorithm
generalized to odd-mass nuclei~\cite{Zhao00}.

In the present review the order of the interactions in the shell-model hamiltonian is limited to two-body
and lowest-order transition operators are taken.

\section{Boson mapping}
\label{s_mapping}
The boson equivalent of the basis~(\ref{e_fnobas}) is
\begin{equation}
|b^nL_2\dots L_n\rangle\equiv
\left(\cdots\left(\left(b^\dag\times b^\dag\right)^{(L_2)}\times b^\dag\right)^{(L_3)}\times\cdots\times
b^\dag\right)^{(L_n)}|{\rm o}\rangle,
\label{e_bnobas}
\end{equation}
where $b^\dag$ creates a boson
with angular momentum (or spin) $\ell=2j$ and isospin $t=0$.
While the angular momentum coupling
is the same in Eqs.~(\ref{e_fnobas}) and~(\ref{e_bnobas}),
overlap and hamiltonian matrix elements are different in both bases
because of the internal structure of the pairs,
in contrast to the assumed elementary character of the bosons.
Nevertheless, Pauli corrections can be systematically applied
to the boson calculation in the following way.
In general, for $n>2$, the boson basis~(\ref{e_bnobas})
is non-orthogonal and overcomplete.
As in the fermion case, the diagonalization of the overlap matrix
\begin{equation}
\langle b^n_i|b^n_{i'}\rangle\equiv
\langle b^nL_2\dots L_{n-1}L_n|b^nL'_2\dots L'_{n-1}L_n\rangle,
\label{e_bover}
\end{equation}
leads to an orthogonal basis of linearly independent vectors.
For technical reasons
that have to do with the computation of coefficients of fractional parentage (CFPs),
it is in this case more convenient to define an orthonormal basis
via a Gram--Schmidt procedure.
For a given sequence of linearly independent, non-orthogonal $n$-boson states
$|b^n_i\rangle$ $(i=1,\dots,\omega')$,
an orthogonal series can be defined as follows:
\begin{eqnarray}
|\tilde b^n_1\rangle&=&\langle b^n_1|b^n_1\rangle^{-1/2}|b^n_1\rangle,
\nonumber\\
|\tilde b^n_2\rangle&=&
\bigl(\langle b^n_2|b^n_2\rangle-\langle b^n_2|\tilde b^n_1\rangle^2\bigr)^{-1/2}
\bigl(|b^n_2\rangle-\langle b^n_2|\tilde b^n_1\rangle|\tilde b^n_1\rangle\bigr),
\nonumber\\ &\vdots&\nonumber\\
|\tilde b^n_i\rangle&=&
\left(\langle b^n_i|b^n_i\rangle-\sum_{j=1}^{i-1}\langle b^n_i|\tilde b^n_j\rangle^2\right)^{-1/2}
\left(|b^n_i\rangle-\sum_{j=1}^{i-1}\langle b^n_i|\tilde b^n_j\rangle|\tilde b^n_j\rangle\right),
\label{e_bobasi}
\end{eqnarray}
until $i=\omega'$.
To establish the connection with the orthogonal fermion-pair series,
an additional transformation is needed,
\begin{equation}
|\bar b^n_k\rangle=\sum_{i=1}^\omega c_{ki}|\tilde b^n_i\rangle,
\qquad
k=1,\dots,\omega,
\label{e_bobas}
\end{equation}
in terms of the coefficients $c_{ki}$ defined in Eq.~(\ref{e_fobas}).
Because of the orthogonality of the basis $|\tilde b^n_i\rangle$
and the properties of the coefficients $c_{ki}$,
the basis $|\bar b^n_k\rangle$ is orthogonal
and is the boson equivalent of the fermion basis $|\bar B^n_k\rangle$.
The matrix elements of the boson hamiltonian in this basis
are therefore determined from
\begin{equation}
\langle\bar b^n_k|\hat H^{\rm b}|\bar b^n_{k'}\rangle=
\langle\bar B^n_k|\hat H^{\rm f}|\bar B^n_{k'}\rangle,
\qquad
k,k'=1,\dots,\omega.
\label{e_bmei}
\end{equation}
With use of the inverse of the relation~(\ref{e_bobas}),
of the equality~(\ref{e_bmei}) and of Eq.~(\ref{e_fme}),
the matrix elements of the boson hamiltonian in the orthogonal basis $|\tilde b^n_i\rangle$
can be written in terms of those
of the shell-model hamiltonian in the fermion-pair basis,
\begin{equation}
\langle\tilde b^n_i|\hat H^{\rm b}|\tilde b^n_{i'}\rangle=
\sum_{k,k'=1}^\omega
\sqrt{\frac{1}{o_ko_{k'}}}c_{ki}c_{k'i'}
\sum_{j,j'=1}^\omega c_{kj}c_{k'j'}
\langle B^n_j|\hat H^{\rm f}|B^n_{j'}\rangle,
\qquad
i,i'=1,\dots,\omega.
\label{e_bme}
\end{equation} 

Three additional technical issues must be ironed out.
First, for a given total angular momentum $L_n$,
the number of linearly independent boson states~(\ref{e_bnobas})
may be larger than the corresponding number of fermion-pair states~(\ref{e_fnobas}),
$\omega\leq\omega'$,
indicating that there are $\omega'-\omega$ spurious boson states
which are Pauli forbidden in the fermion space.
The matrix elements of the boson hamiltonian
pertaining to these states remain undefined in Eq.~(\ref{e_bme}).
Since these states are spurious,
they must be eliminated from the boson space,
implying the following choice of boson matrix elements:
\begin{eqnarray}
\langle\tilde b^n_i|\hat H^{\rm b}|\tilde b^n_i\rangle&=&+\infty,
\qquad
i=\omega+1,\dots,\omega',
\nonumber\\
\langle\tilde b^n_i|\hat H^{\rm b}|\tilde b^n_{i'}\rangle=
\langle\tilde b^n_{i'}|\hat H^{\rm b}|\tilde b^n_i\rangle&=&0,
\qquad
i\leq \omega<i'\leq \omega'.
\label{e_bmee}
\end{eqnarray}

The second technical issue concerns the fact that
Eq.~(\ref{e_bme}) defines the entire boson hamiltonian
up to and including $n$-body interactions.
To isolate its $n$-body part $\hat H^{\rm b}_n$,
one should subtract the previously determined $n'$-body interactions, $n'<n$.
The procedure is straightforward
but rather cumbersome to write down explicitly up to all orders.
Up to the three-body interactions that will be considered below,
one has the following results.
The single-boson energy is determined from
\begin{equation}
\epsilon_b\equiv
\langle b|\hat H^{\rm b}|b\rangle=
\langle B|\hat H^{\rm f}|B\rangle,
\label{e_bme1}
\end{equation}
which is nothing but the shell-model matrix element $\upsilon^{2\rm f}_{JT}$
in the aligned neutron--proton configuration with $J=2j$ and $T=0$.
The two-body part of the boson hamiltonian is determined from
\begin{equation}
\upsilon^{2\rm b}_{L_2}\equiv
\langle b^2L_2|\hat H^{\rm b}_2|b^2L_2\rangle=
\langle b^2L_2|\hat H^{\rm b}|b^2L_2\rangle-2\epsilon_b,
\label{e_bme2}
\end{equation}
where it is assumed that the two-boson states are normalized
such that the matrix element of the total boson hamiltonian
can be taken from Eq.~(\ref{e_bme}).
The three-body part of the boson hamiltonian follows from
\begin{eqnarray}
\langle b^3[\tilde L_2]L_3|\hat H^{\rm b}_3|b^3[\tilde L'_2]L_3\rangle&=&
\langle b^3[\tilde L_2]L_3|\hat H^{\rm b}|b^3[\tilde L'_2]L_3\rangle
-3\epsilon_b\delta_{\tilde L_2\tilde L'_2}-
\nonumber\\&&
3\sum_{L''_2}
[\ell^2[L''_2]\ell|\}\ell^3[\tilde L_2]L_3]
[\ell^2[L''_2]\ell|\}\ell^3[\tilde L'_2]L_3]
\upsilon^{2\rm b}_{L''_2},
\label{e_bme3}
\end{eqnarray}
where again the matrix element of the total boson hamiltonian on the right-hand side
are calculated from Eq.~(\ref{e_bme}).
Equation~(\ref{e_bme3}) requires some explanation.
The basis consisting of the three-boson states
\begin{equation}
|b^3L_2L_3\rangle\equiv
\left(\left(b^\dag\times b^\dag\right)^{(L_2)}\times b^\dag\right)^{(L_3)}|{\rm o}\rangle,
\label{e_bnobas3}
\end{equation}
is non-orthogonal and non-normalized.
The intermediate angular momentum $L_2$ can be used as a label
and, after the application of Eq.~(\ref{e_bobasi}),
one arrives at an orthogonal basis denoted by $|b^3[\tilde L_2]L_3\rangle$,
with the notation $[\tilde L_2]$ as a reminder of the Gram--Schmidt procedure.
This basis can be used to express the matrix elements of a two-body interaction
in the usual manner with CFPs~\cite{Talmi93}.
In general, the matrix element of an $n'$-body boson hamiltonian
between $n$-boson states ($n\geq n'$) can be written as
\begin{equation}
\langle\tilde b^n_i|\hat H^{\rm b}_{n'}|\tilde b^n_{i'}\rangle=
\frac{n!}{n'!(n-n')!}
\sum_{jkk'}
[\tilde b^{n-n'}_j\tilde b^{n'}_k|\}\tilde b^n_i]
[\tilde b^{n-n'}_j\tilde b^{n'}_{k'}|\}\tilde b^n_{i'}]
\langle\tilde b^{n'}_k|\hat H^{\rm b}_{n'}|\tilde b^{n'}_{k'}\rangle,
\label{e_bmen}
\end{equation} 
in terms of $n\rightarrow n-n'$ CFPs.
The third term on the right-hand side of Eq.~(\ref{e_bme3})
arises from the application of this result for $n=3$ and $n'=2$,
together with the explicit notation of CFPs for bosons with spin $\ell$.

The third technical issue concerns the hierarchy of states
since, in general, the definition of the interactions between the bosons
depends on the order of states
chosen in the Gram--Schmidt procedure~(\ref{e_bobas}).
In the mapping from $B$ pairs to $b$ bosons
no ambiguity exists for the two-body interaction between the bosons ($n=2$)
since states are unique for a given angular momentum $J$.
This is no longer the case for $n\geq3$
and as a result there exist many different $n$-body interactions
that exactly reproduce the shell-model results in the $B^n$ space.
The method followed here is to define a hierarchy
based on the importance of the overlap with the yrast shell-model state
(see Sect.~\ref{s_appli} for examples),
leading to a procedure which, for $n=3$, can be summarized in the following steps.
\begin{itemize}
\item
Construct and diagonalize the shell-model hamiltonian in the $B^3$ basis,
leading to the eigenvalues $E_k$ ($k=1,\dots,\omega$).
\item
To deal with spurious states,
the set of $\omega$ eigenvalues
is enlarged with $E_k=+\infty$ ($k=\omega'-\omega+1,\dots,\omega'$)
({\it i.e.}, $\omega'-\omega$ large values in numerical applications).
\item
Construct and diagonalize the boson hamiltonian
with up to two-body interactions in the analogue $b^3$ basis.
This is achieved by computing the second and third terms
on the left-hand side of Eq.~(\ref{e_bme3})
which after diagonalization yields the eigenvalues $E'_k$ ($k=1,\dots,\omega'$)
with corresponding eigenvectors $c'_{kl}$ ($l=1,\dots,\omega'$).
\item
The three-body interaction in the analogue boson basis
is obtained by transforming back the matrix
with the differences $E_k-E'_k$ on the diagonal,
\begin{equation}
\upsilon^{3\rm b}_{ll'}=
\sum_{k=1}^{\omega'}c'_{lk}(E_k-E'_k)c'_{kl'},
\label{e_bme3b}
\end{equation}
where $l$ and $l'$ are short-hand notations
for the three-boson labels
$[\tilde L_2]L_3$ and $[\tilde L'_2]L_3$.
\end{itemize}

Since the three-body components of the boson interaction
are found to be small (see Sect.~\ref{s_appli}),
no exhaustive study of the three-body aspect of the mapping
is attempted in this review.

\section{Three approximations}
\label{s_approx}
Let us now take stock of the situation with regard to the aligned-pair approximation
as described in the technical discussion of the previous two sections.
A first possibility is to determine the $T=0$ spectrum of a $2n$-particle system
by diagonalizing a given shell-model hamiltonian $\hat H^{\rm f}$
in the aligned-pair basis $|\bar B^n_k\rangle$.
This is a truncated shell-model calculation
in which the Pauli principle is fully taken into account
and no boson mapping is needed.
The calculation becomes more difficult as $n$ increases
because of the complexity of Chen's algorithm.
This truncated shell-model calculation
can be replaced {\em exactly} by its boson equivalent
{\em if} the mapped boson hamiltonian $\hat H^{\rm b}$
is determined up to all orders
({\it i.e.}, up to order $n$ for a $2n$-particle system).
The Pauli principle is obeyed
by means of appropriate interactions between the bosons.
No simplification of the original problem is obtained
since the determination of $\hat H^{\rm b}$ up to order $n$
requires the calculation of matrix elements of $\hat H^{\rm f}$
in the aligned-pair basis $|B^n_k\rangle$
[see Eq.~(\ref{e_bme})].
Significant simplifications may result, however,
if the mapped boson hamiltonian $\hat H^{\rm b}$
is determined up to an order $n'<n$
but this simplification is at the expense of some violation of the Pauli principle.

In Sect.~\ref{s_appli} the above statements are illustrated with examples.
Since a number of approximations are made at different stages,
it is useful to enounce these approximations
and to indicate whether they can be tested theoretically and/or experimentally.
Let us start from the general observation
that the $N=Z$ nuclei under consideration
can be described in the spherical shell model
if a sufficiently large model space with an appropriate interaction is adopted.
With this as a premise the following assumptions are made
to arrive at an approximation in terms of aligned-pair bosons.
\begin{romanlist}[(iii)]
\item
{\it The shell-model space is truncated to a single high-$j$ orbit.}
A theoretical test of this assumption is not attempted in this review.
Its validity clearly depends on the specific features of the initial shell-model hamiltonian.
Two particular mass regions where the approximation might be valid spring to mind:
$N=Z$ nuclei in the $1f_{7/2}$ and $1g_{9/2}$ shells.
More important is that the assumption can be tested experimentally,
as illustrated with examples in Sect.~\ref{s_appli}.
\item
{\it The single-$j$ shell space is reduced to one written in terms of aligned $B$ pairs.}
Some dependence exists on the shell-model interaction
adopted in the single-$j$ shell space.
Nevertheless, if a reasonable interaction is taken,
this assumption can be tested generically.
Furthermore, the extension of the NPSM that includes isospin~\cite{Fu13a}
is the appropriate formalism to test the combined approximations (i) and (ii).
A recent calculation of this type~\cite{Fu13b},
which starts from a realistic shell-model space and interaction, 
seem to indicate that the combined approximations (i) and (ii)
hold fairly well in $N=Z$ nuclei from $^{92}$Pd to $^{100}$Sn.
\item
{\it The aligned $B$ pairs are replaced by $b$ bosons.}
As argued in the previous section,
if the boson hamiltonian is calculated up to all orders,
the mapping is exact and no approximation is made.
The usual procedure, however, is to map up to two-body boson interactions
which implies some amount of Pauli violation.
In the next section the validity of the two-body boson mapping is tested
by calculating the effect of the three-body interaction.
\end{romanlist}

\section{Applications}
\label{s_appli}
A number of results can be established for a shell with arbitrary $j$ .
They are useful in the discussion of specific cases,
in particular the $1f_{7/2}$ and $1g_{9/2}$ shells.

\subsection{Any $j$ shell}
\label{ss_j}
The M1 operator in the shell model is given by
\begin{equation}
\hat T^{\rm f}_\mu({\rm M1})=
\sqrt{\frac{3}{4\pi}}
\left(\sum_{i\in\nu}g^s_\nu s_\mu(i)+
\sum_{i\in\pi}\left[g^\ell_\pi \ell_\mu(i)+g^s_\pi s_\mu(i)\right]\right),
\label{e_fm1a}
\end{equation}
where the sums are over neutrons and protons,
and in each sum appear the orbital and spin gyromagnetic factors, $g^\ell_\rho$ and $g^s_\rho$,
with $\rho=\nu$ for a neutron and $\rho=\pi$ for a proton.
For the calculation of magnetic moments ({\it i.e.}, diagonal matrix elements)
the M1 operator~(\ref{e_fm1a}) can be replaced by one
in terms of neutron and proton $g$ factors.
In second quantization the $z$ component of the latter operator
can be written as
\begin{equation}
\hat \mu^{\rm f}_0=
\sqrt{\frac{j(j+1)(2j+1)}{3}}
\left[g_\nu(\nu^\dag_j\times\tilde\nu_j)^{(1)}_0+
g_\pi(\pi^\dag_j\times\tilde\pi_j)^{(1)}_0\right],
\label{e_fm1b}
\end{equation}
where $\rho^\dag_{jm}$ creates a neutron ($\rho=\nu$) or a proton ($\rho=\pi$) in the $j$ shell,
and $\tilde\rho_{jm}=(-)^{j+m}\rho_{j-m}$.
This operator can be written alternatively
as a sum of an isoscalar part, multiplied by $(g_\nu+g_\pi)$,
and an isovector part, multiplied by $(g_\nu-g_\pi)$.
For the M1 matrix elements between states in a single-$j$ shell
of the same isospin $T$ and with projection $T_z=0$,
only the former part contributes
and, since the isoscalar part is proportional to the angular momentum operator,
it follows that the $g$ factor of any state in an $N=Z$ nucleus
equals $(g_\nu+g_\pi)/2$.
This result is generally valid under the assumptions
that isospin is a good quantum number
and that the nucleons are confined to a single-$j$ shell~\cite{Lawson71}.

In terms of $b$ bosons the M1 operator is of the form
\begin{equation}
\hat T^{\rm b}_\mu({\rm M1})=
\sqrt{\frac{3}{4\pi}}g_b\hat J_\mu=
\sqrt{\frac{2j(2j+1)(4j+1)}{4\pi}}g_b
(b^\dag\times\tilde b)^{(1)}_\mu.
\label{e_bm1}
\end{equation}
The $g$ factor of the $b$ boson, $g_b$,
is obtained from the $g$ factor of the $B$ pair
which, due to the above argument, equals $(g_\nu+g_\pi)/2$.
Since the operator~(\ref{e_bm1}) is proportional to the angular momentum operator,
one finds that the $g$ factor of any state $|\alpha J\rangle$ in the boson model equals
\begin{equation}
g(\alpha J)\equiv
\frac{\mu(\alpha J)}{J}=
\sqrt{\frac{4\pi}{3}}
\frac{\langle\alpha JJ|\hat T^{\rm b}_0({\rm M1})|\alpha JJ\rangle}{J}=
g_b=\frac{g_\nu+g_\pi}{2}.
\label{e_bg}
\end{equation}
One recovers therefore the shell-model result
that the $g$ factor of any state in an $N=Z$ nucleus equals $(g_\nu+g_\pi)/2$.

The conclusion of the preceding discussion
is that magnetic moments do not provide a test
of the assumptions (ii) and (iii) of Sect.~\ref{s_approx}
since any $T=T_z=0$ state in a single-$j$ shell 
has a $g$ factor equal to $(g_\nu+g_\pi)/2$,
irrespective of whether this state can be written in terms of $B$ pairs or not,
and since the same result is obtained with $b$ bosons.
However, deviations from $(g_\nu+g_\pi)/2$
are indicative of admixtures of configurations beyond a single-$j$ shell
and therefore magnetic moments constitute a test of assumption (i).

The E2 operator in the shell model is
\begin{equation}
\hat T^{\rm f}_\mu({\rm E2})=
e_\nu\sum_{i\in\nu}r_i^2 Y_{2\mu}(\theta_i,\phi_i)+
e_\pi\sum_{i\in\pi}r_i^2 Y_{2\mu}(\theta_i,\phi_i),
\label{e_fe2a}
\end{equation}
where each sum is multiplied with the appropriate effective charge.
In a single-$j$ shell the second-quantized form of this operator is
\begin{equation}
\hat T^{\rm f}_\mu({\rm E}2)=
-x_j\left(N+\textstyle{{\frac32}}\right)l_{\rm ho}^2
\left[e_\nu(\nu^\dag_j\times\tilde\nu_j)^{(2)}_\mu+
e_\pi(\pi^\dag_j\times\tilde\pi_j)^{(2)}_\mu\right],
\label{e_fe2b}
\end{equation}
where $N$ is the major oscillator quantum number [$N=2(n-1)+\ell$]
and $l_{\rm ho}$ is the length parameter of the harmonic oscillator,
and with
\begin{equation}
x_j=
\left[\frac{(2j-1)(2j+1)(2j+3)}{64\pi j(j+1)}\right]^{1/2}.
\label{e_xj}
\end{equation}

In terms of $b$ bosons the E2 operator is of the form
\begin{equation}
\hat T^{\rm b}_\mu({\rm E2})=
e_b(b^\dag\times\tilde b)^{(2)}_\mu,
\label{e_be2}
\end{equation}
where the boson effective charge $e_b$
is obtained from the condition
\begin{equation}
\langle B||\hat T^{\rm f}({\rm E2})||B\rangle=
\langle b||\hat T^{\rm b}({\rm E2})||b\rangle.
\label{e_bfe2}
\end{equation}
The two-particle matrix element on the left-hand side of Eq.~(\ref{e_bfe2})
can be readily derived with the help of Eq.~(\ref{e_fe2b}),
leading to the result
\begin{equation}
e_b=-(e_\nu+e_\pi)
\left(N+\textstyle{{\frac32}}\right)l_{\rm ho}^2
\left[\frac{(2j-1)^2(2j+1)(4j+1)(4j+3)}{128\pi j(j+1)^2(4j-1)}\right]^{1/2}.
\label{e_eb}
\end{equation}
Unlike the case of M1 properties,
no general conclusions can be drawn for E2 transitions and moments.
The preceding expressions are nevertheless helpful
in the discussion of nuclei in the two shells of interest. 

\subsection{The $1f_{7/2}$ shell}
\label{ss_f7}
The restriction to a single-$j$ shell is an approximation
which, if valid at all, induces higher-order interactions
in the effective shell-model hamiltonian
that should be calculated from perturbation theory~\cite{Poves81}.
To avoid the complexities
associated with three- and higher-body interactions between nucleons,
a more phenomenological approach is followed here
which consists of introducing a  two-body interaction
that depends on the mass number $A$.
The spectra of the nuclei $^{42}$Sc and $^{54}$Co are well known~\cite{NNDC}
and allow the determination of the particle--particle
and hole--hole matrix elements, respectively, up to a constant.
This constant is determined from measured binding energies~\cite{Audi12}
of neighbouring nuclei,
leading to the shell-model matrix elements shown in Table~\ref{t_fint}.
\begin{table}[pt]
\tbl{\label{t_fint}
Shell-model matrix elements $\upsilon^{2\rm f}_{JT}$ (in MeV) in the $1f_{7/2}$ and $1g_{9/2}$ shells.}
{\begin{tabular}{@{}ccccccccccc@{}}
\toprule
$(JT)$&(01)&(10)&(21)&(30)&(41)&(50)&(61)&(70)&(81)&(90)\\
\colrule
$^{42}$Sc&
$-3.187$&$-2.576$&$-1.601$&$-1.697$&$-0.372$&$-1.677$&$0.055$&$-2.571$&&\\
$^{54}$Co&
$-2.551$&$-1.614$&$-1.105$&$-0.730$&$\hphantom{-}0.101$&$-0.664$&$0.349$&$-2.354$&&\\
\colrule
SLGT0&
$-2.392$&$-1.546$&$-0.906$&$-0.747$&$-0.106$&$-0.423$&
$0.190$&$-0.648$&$0.321$&$-1.504$\\
\botrule
\end{tabular}}
\end{table}
The interaction appropriate for $N=Z$ nuclei intermediate between $^{42}$Sc and $^{54}$Co
is obtained from linear interpolation,
\begin{equation}
\upsilon^{2\rm f}_{JT}(A)=
\frac{54-A}{12}\upsilon^{2\rm f}_{JT}(^{42}{\rm Sc})+
\frac{A-42}{12}\upsilon^{2\rm f}_{JT}(^{54}{\rm Co}).
\label{e_vf}
\end{equation}

The advantage of using a two-body interaction
is that particle--hole symmetry is preserved.
The calculation of a nucleus heavier than $^{48}$Cr,
corresponding to the space $(1f_{7/2})^{2n}$ with $n>4$,
can be replaced by one in the space $(1f_{7/2})^{16-2n}$.
All properties are identical except quadrupole moments
which change sign~\cite{Lawson80}.
This simplifies the calculation in the pair basis $|P^n_j\rangle$
which for $n>4$ can be replaced by $|P^{8-n}_j\rangle$.

\subsubsection{$^{44}$Ti and $^{52}$Fe}
\label{sss_tife}
\begin{figure}
\includegraphics[height=4cm]{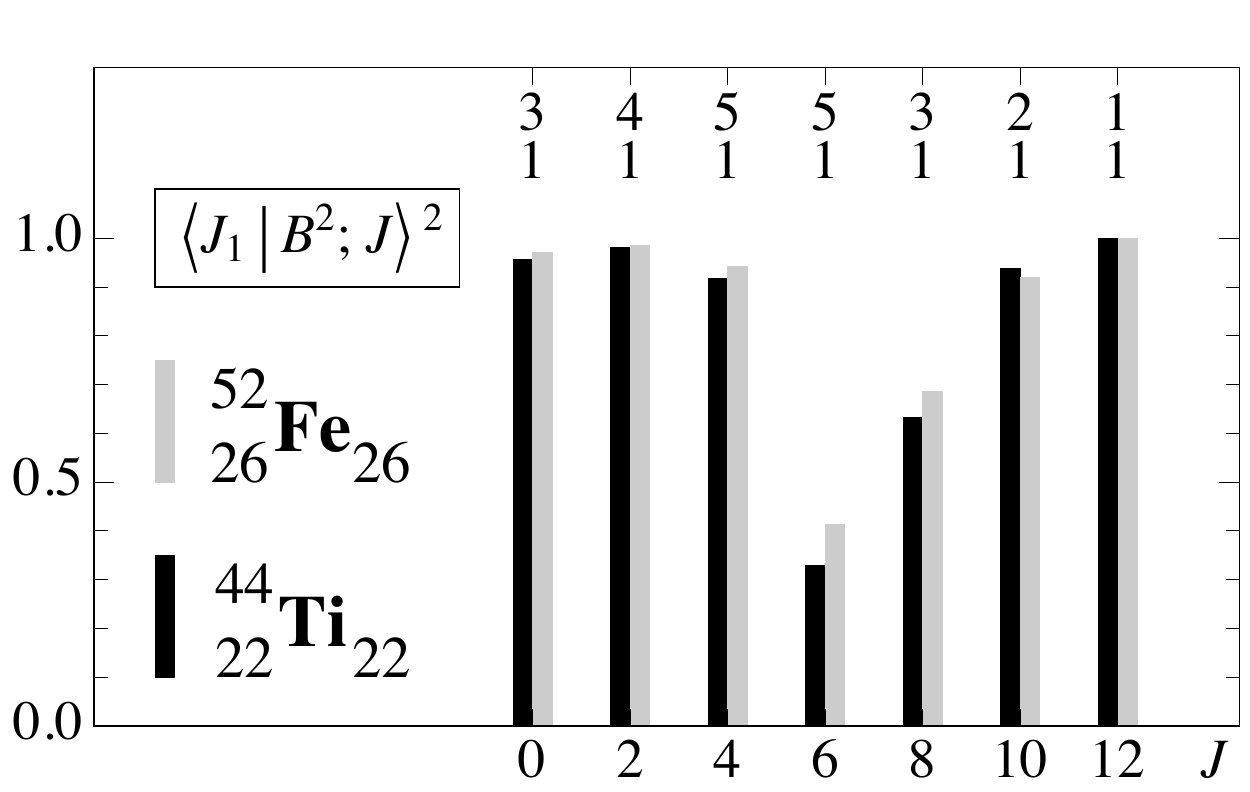}
\caption{
\label{f_bb_tife}
Overlaps of the yrast eigenstates in the $(1f_{7/2})^4$ system,
for angular momentum $J$ and isospin $T=0$,
with the $B$-pair state $|B^2;J\rangle$.
The shell-model interaction is defined in Eq.~(\ref{e_vf}).
Also shown are the numbers of $(1f_{7/2})^4$ states (top)
and of $B$-pair states (bottom)
with angular momentum $J$ and isospin $T=0$.}
\end{figure}
For two neutrons and two protons (both particle- or hole-like)
the $B$-pair state is unique for a given total angular momentum $J$ and isospin $T=0$.
The $B$-pair content of a given shell-model state
can be obtained from Eq.~(\ref{e_fana}) with $\omega=1$.
This quantity is shown in Fig.~\ref{f_bb_tife} for the yrast states in $^{44}$Ti and $^{52}$Fe.
Most yrast states have a large $B$-pair content
but not for $J=6$ and $J=8$.
It seems as if the two $B$ pairs do not like to couple
to a total angular momentum which equals their individual spins.
Although the interaction varies considerably with mass (see Table~\ref{t_fint}),
similar results are found in $^{44}$Ti and $^{52}$Fe,
indicating that these conclusions are robust
as long as a reasonable nuclear interaction is used.

Also shown in Fig.~\ref{f_bb_tife}
are the numbers of $(1f_{7/2})^4$ states and of $B$-pair states
with angular momentum $J$ and isospin $T=0$.
This allows one to judge
whether the observation of a high overlap is trivial or meaningful.
For example, only one shell-model state exists with $J=12$ and $T=0$
which therefore necessarily has an overlap of 1 with the $B$-pair state.
In contrast, four shell-model states exist with $J=2$ and $T=0$
but it is found that the yrast eigenstate
has an overlap of more than 0.98 with a single $B$-pair state.
The latter is a physically meaningful result
whereas the former is trivial.

\begin{figure}
\includegraphics[width=8cm]{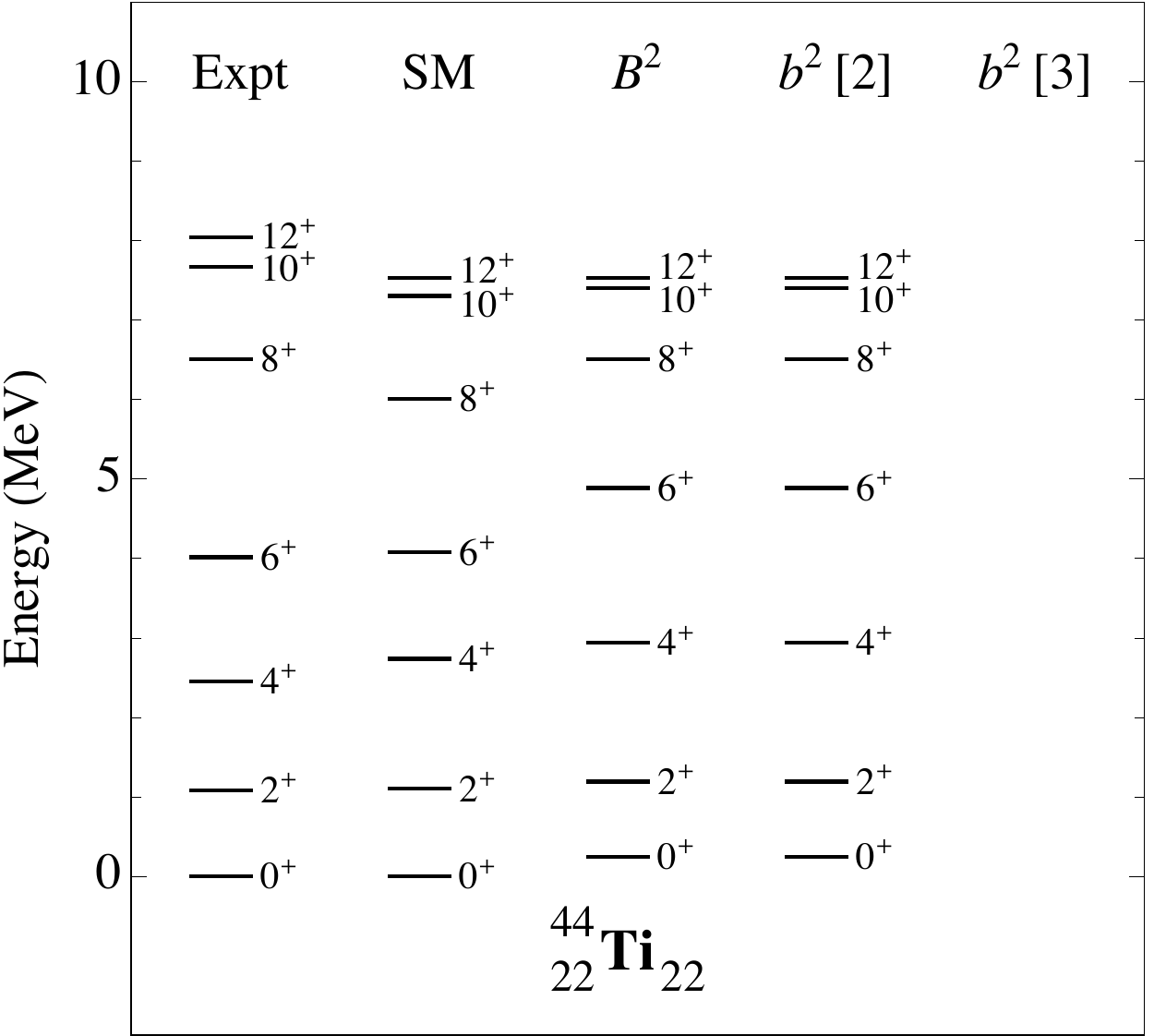}
\caption{
\label{f_spec44}
The yrast spectrum of $^{44}$Ti.
Levels are labelled by their angular momentum and parity $J^\pi$.
The different columns contain the experimental~\cite{NNDC} levels (Expt),
the results of the $(1f_{7/2})^4$ shell model (SM) with the interaction~(\ref{e_vf}),
the expectation value of the shell-model hamiltonian in the $B$-pair state ($B^2$)
and the expectation value of the mapped boson hamiltonian
with up to two-body interactions ($b^2[2]$).
The shell-model energy of the $0^+_1$ level is normalized to zero.}
\end{figure}
\begin{figure}
\includegraphics[width=8cm]{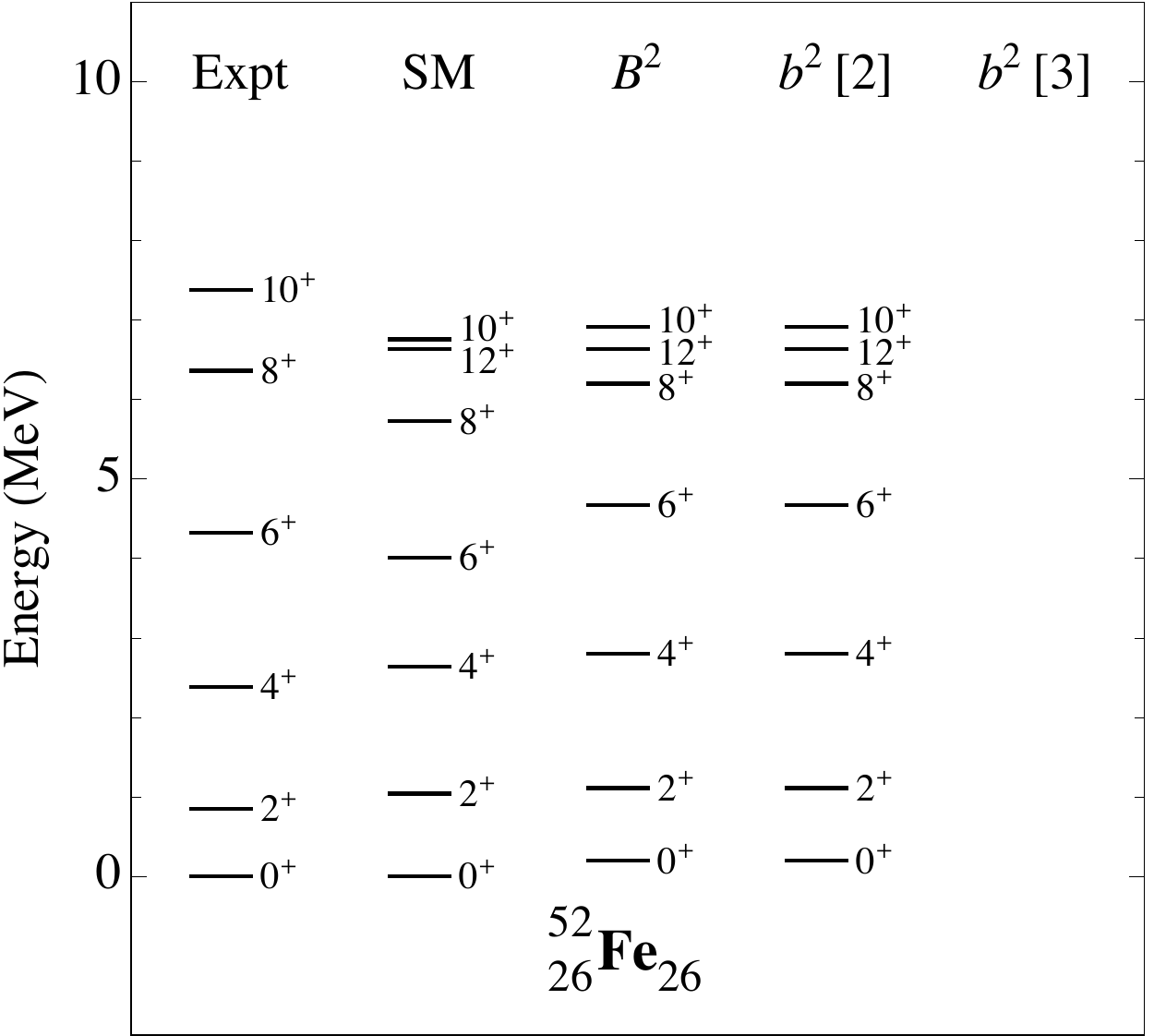}
\caption{
\label{f_spec52}
Same as Fig.~\ref{f_spec44} for $^{52}$Fe.}
\end{figure}
The energy spectra of $^{44}$Ti and $^{52}$Fe,
shown in Figs.~\ref{f_spec44} and~\ref{f_spec52},
confirm the above wave-function analysis.
For the sake of comparison with the data,
the shell-model energy of the $0^+$ level is normalized to zero,
and it is seen that the excitation spectra
calculated in the shell model (SM)
are reasonably close to the observed ones.
The column `$B^2$' shows the expectation value
of the shell-model hamiltonian in the $B$-pair state $|B^2;J,T=0\rangle$.
Note that absolute energies are calculated
which are plotted relative to the shell-model $0^+_1$ level.  
Therefore, the differences in energy between corresponding levels
in the `SM' and `$B^2$' columns
correlate with the overlaps shown in Fig.~\ref{f_bb_tife}.
For example, the difference is greatest for $J^\pi=6^+$
since for this state the overlap is smallest.

\begin{table}[pt]
\tbl{\label{t_bmef}
Coefficients $a^L_{JT}(j)$ in the expansion~(\ref{e_bmej}) for $j=7/2$.}
{\begin{tabular}{@{}ccccccccc@{}}
\toprule
&\hspace{-15pt}$L$\hspace{-15pt}&0&2&4&6&8&10&12\\[-2pt]
$(JT)$&&&&&&&&\\[-2pt]
\colrule
$(01)$&&
$\frac{19305}{13732}$&
$\phantom{-}\frac{1287}{1312}$&
$\frac{3315}{8612}$&
$\frac{14535}{245288}$&&&\\[3pt]
$(10)$&&
$\frac{35035}{41196}$&
$\phantom{-}\frac{2821}{3936}$&
$\frac{11305}{25836}$&
$\frac{110789}{735864}$&
$\frac{38}{2913}$&&\\[3pt]
$(21)$&&
$\frac{21021}{13732}$&
$\phantom{-}\frac{2379}{1312}$&
$\frac{16711}{8612}$&
$\frac{332367}{245288}$&
$\frac{399}{971}$&&\\[3pt]
$(30)$&&
$\frac{22295}{151052}$&
$\phantom{-}\frac{3969}{14432}$&
$\frac{145775}{284196}$&
$\frac{1779141}{2698168}$&
$\frac{5047}{10681}$&
$\frac{115}{1221}$&\\[3pt]
$(41)$&&
$\frac{9555}{151052}$&
$\phantom{-}\frac{2949}{14432}$&
$\frac{62475}{94732}$&
$\frac{4007955}{2698168}$&
$\frac{45465}{21362}$&
$\frac{2415}{1628}$&\\[3pt]
$(50)$&&
$\frac{245}{178516}$&
$\phantom{-}\frac{147}{17056}$&
$\frac{16415}{335868}$&
$\frac{46305}{245288}$&
$\frac{12515}{25246}$&
$\frac{4711}{5772}$&
$\frac{6}{13}$\\[3pt]
$(61)$&&
$\frac{15}{151052}$&
$\phantom{-}\frac{21}{14432}$&
$\frac{1435}{94732}$&
$\frac{270627}{2698168}$&
$\frac{9843}{21362}$&
$\frac{2469}{1628}$&
$3$\\[3pt]
$(70)$&&
$\frac{1}{1716}$&
$-\frac{197}{562848}$&
$\frac{2293}{3694548}$&
$\frac{10337}{8094504}$&
$\frac{15589}{833118}$&
$\frac{1897}{21164}$&
$\frac{7}{13}$\\[3pt]
\botrule
\end{tabular}}
\end{table}
\begin{table}[pt]
\tbl{\label{t_bint}
Boson interaction matrix elements $\upsilon^{2\rm b}_L$ (in MeV)
appropriate for the $1f_{7/2}$ and $1g_{9/2}$ shells.}
{\begin{tabular}{@{}ccccccccccc@{}}
\toprule
$L$&0&2&4&6&8&10&12&14&16&18\\
\colrule
$^{42}$Sc&
$-3.187$&$-2.576$&$-1.601$&$-1.697$&$-0.372$&$-1.677$&$0.055$&$+\infty$&&\\
$^{54}$Co&
$-2.551$&$-1.614$&$-1.105$&$-0.730$&$0.101$&$-0.664$&$0.349$&$+\infty$&&\\
\colrule
SLGT0&
$-5.635$&$-4.956$&$-3.694$&$-2.333$&$-1.209$&$-0.455$&
$-0.062$&$0.058$&$-0.506$&$+\infty$\\
\botrule
\end{tabular}}
\end{table}
The two-boson calculation with up to two-body interactions,
shown in the column `$b^2[2]$' of Fig.~\ref{f_bb_tife},
reproduces exactly the $B$-pair calculation,
in agreement with the discussion of Sect.~\ref{s_mapping}.
Since, for a given angular momentum $J$ and isospin $T=0$,
the mapping from two $B$ pairs to two $b$ bosons is one-to-one,
simple expressions are found
for the boson interaction matrix elements $\upsilon^{2\rm b}_L$
in terms of the two-body fermion matrix elements $\upsilon^{2\rm f}_{JT}$.
These relations are of the generic form
\begin{equation}
\upsilon^{2\rm b}_L=
\sum_{JT}a^L_{JT}(j)\upsilon^{2\rm f}_{JT},
\label{e_bmej}
\end{equation}
with coefficients $a^L_{JT}(j)$ that depend
on the single-particle angular momentum $j$ of the shell.
The coefficients for $j=7/2$ are given in Table~\ref{t_bmef}
and the resulting boson interaction matrix elements $\upsilon^{2\rm b}_L$
in Table~\ref{t_bint}.
There is no four-particle shell-model state with $J=14$,
implying the choice $\upsilon^{2\rm b}_{14}=+\infty$,
in line with the recipe~(\ref{e_bmee}).
In numerical calculations a large repulsive matrix element is taken.

No three-body interactions between the bosons
intervene in $^{44}$Ti and $^{52}$Fe.

\subsubsection{$^{46}$V and $^{50}$Mn}
\label{sss_vmn}
\begin{figure}
\includegraphics[height=4cm]{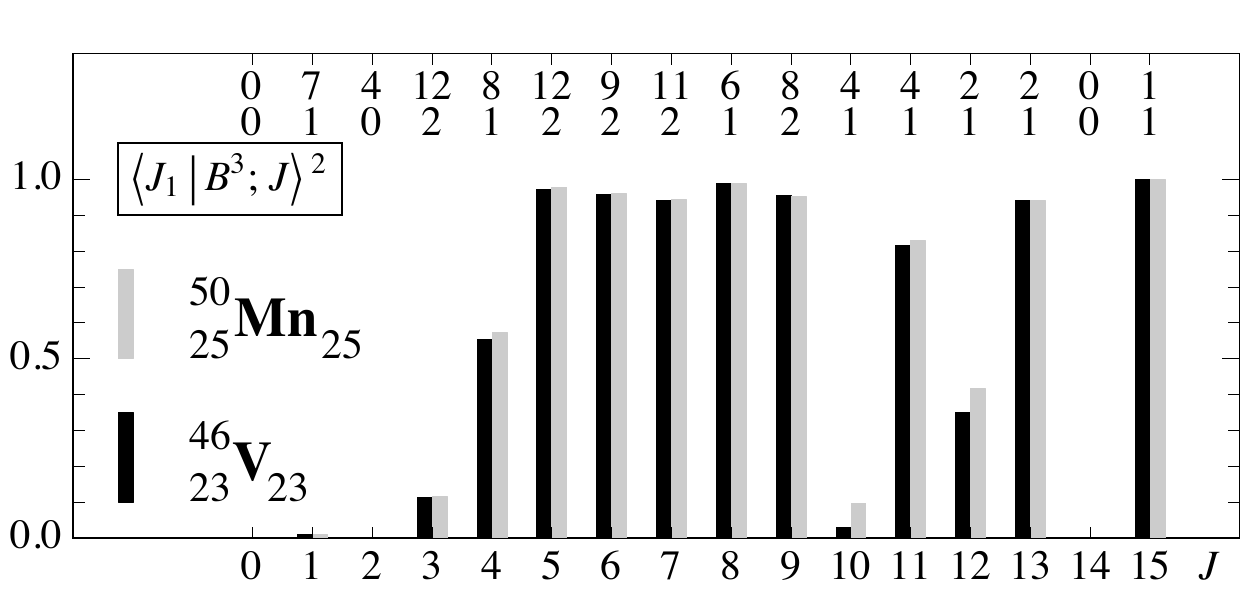}
\caption{
\label{f_bb_vmn}
The square of the projection of the yrast eigenstates in the $(1f_{7/2})^6$ system
onto the subspace spanned by the $B$-pair states $|B^3;J\rangle$,
for angular momentum $J$ and isospin $T=0$.
The shell-model interaction is defined in Eq.~(\ref{e_vf}).
Also shown are the numbers of $(1f_{7/2})^6$ states (top)
and of $B$-pair states (bottom)
with angular momentum $J$ and isospin $T=0$.}
\end{figure}
Odd--odd $N=Z$ nuclei are of particular interest
with regard to the question of the relevance of neutron--proton pairs.
For three neutrons and three protons (both particle- or hole-like) in a $j=7/2$ shell,
there are at most two linearly independent $B$-pair states
for a given total angular momentum $J$ and isospin $T=0$.
The $B$-pair content of a shell-model state
can therefore be obtained from Eq.~(\ref{e_fana}) with $\omega=1$ or 2.
This quantity is shown in Fig.~\ref{f_bb_vmn}
for yrast states in $^{46}$V and $^{50}$Mn.
On top of the figure are shown
the numbers of $(1f_{7/2})^6$ states and of $B$-pair states
with angular momentum $J$ and isospin $T=0$,
in order to judge whether a large overlap
is a physically meaningful or a trivial result.

A surprising feature of the results of Fig.~\ref{f_bb_vmn}
is the `schizophrenic' behaviour of $T=0$ states in $^{46}$V and $^{50}$Mn,
with most having either a large or a small $B$-pair component.
Clearly, only the former states can be interpreted in terms of $B$ pairs or $b$ bosons,
as will be shown below.
Before doing so, a few words are in order
about those states that do {\em not} conform to such a description.
The most obvious example is the $J=2$ state
which simply cannot be constructed out of three $B$ pairs.
A wave-function analysis with the method outlined in Sect.~\ref{s_npsm},
gives $|SPD;2\rangle$ as its main component,
$\langle2^+_1|SPD;2\rangle^2=0.825~(0.841)$ in $^{46}$V ($^{50}$Mn),
where $S$, $P$ and $D$ are pairs
with $J=0,T=1$, $J=1,T=0$ and $J=2,T=1$, respectively.
All low-spin states can in fact be adequately written
in terms of the $S$, $P$, $D$ and $F$ pairs
that correspond to the bosons of \mbox{IBM-4},
confirming the analysis of Juillet {\it et al.}~\cite{Juillet01} in a different mass region.
The most remarkable state of this kind is the yrast $1^+$ level
which approximately can be written as $|P^3;1\rangle$
since $\langle1^+_1|P^3;1\rangle^2=0.725~(0.728)$ in $^{46}$V ($^{50}$Mn).
(Note that there is only one $P^3$ state with angular momentum $J=1$
since $|P^3[0]1\rangle\propto|P^3[2]1\rangle$.)

\begin{figure}
\includegraphics[width=8cm]{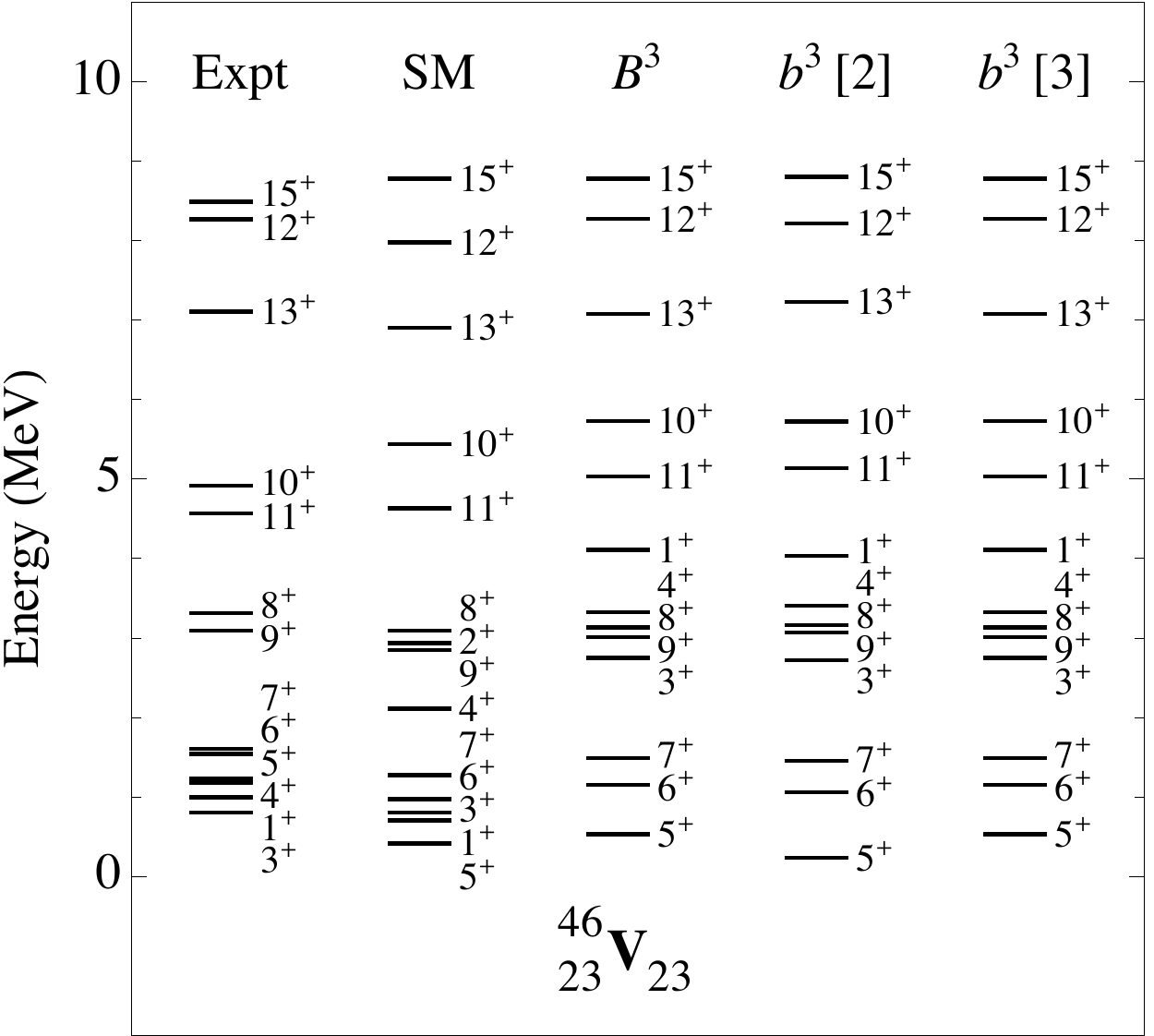}
\caption{
\label{f_spec46}
The spectrum of yrast states in $^{46}$V with isospin $T=0$.
Levels are labelled by their angular momentum and parity $J^\pi$.
The different columns contain the experimental~\cite{NNDC} levels (Expt),
the results of the $(1f_{7/2})^6$ shell model (SM) with the interaction~(\ref{e_vf}),
the lowest eigenvalue of the shell-model hamiltonian in the $B$-pair subspace ($B^3$)
and the lowest eigenvalue of the mapped boson hamiltonian
with up to two-body ($b^3[2]$) and up to three-body ($b^3[3]$) interactions.
The shell-model energy of the $T=0$ ground state, $J^\pi=3^+$,
is normalized to the experimental excitation energy of this level
which is relative to the $0^+$ ground state with isospin $T=1$.}
\end{figure}
\begin{figure}
\includegraphics[width=8cm]{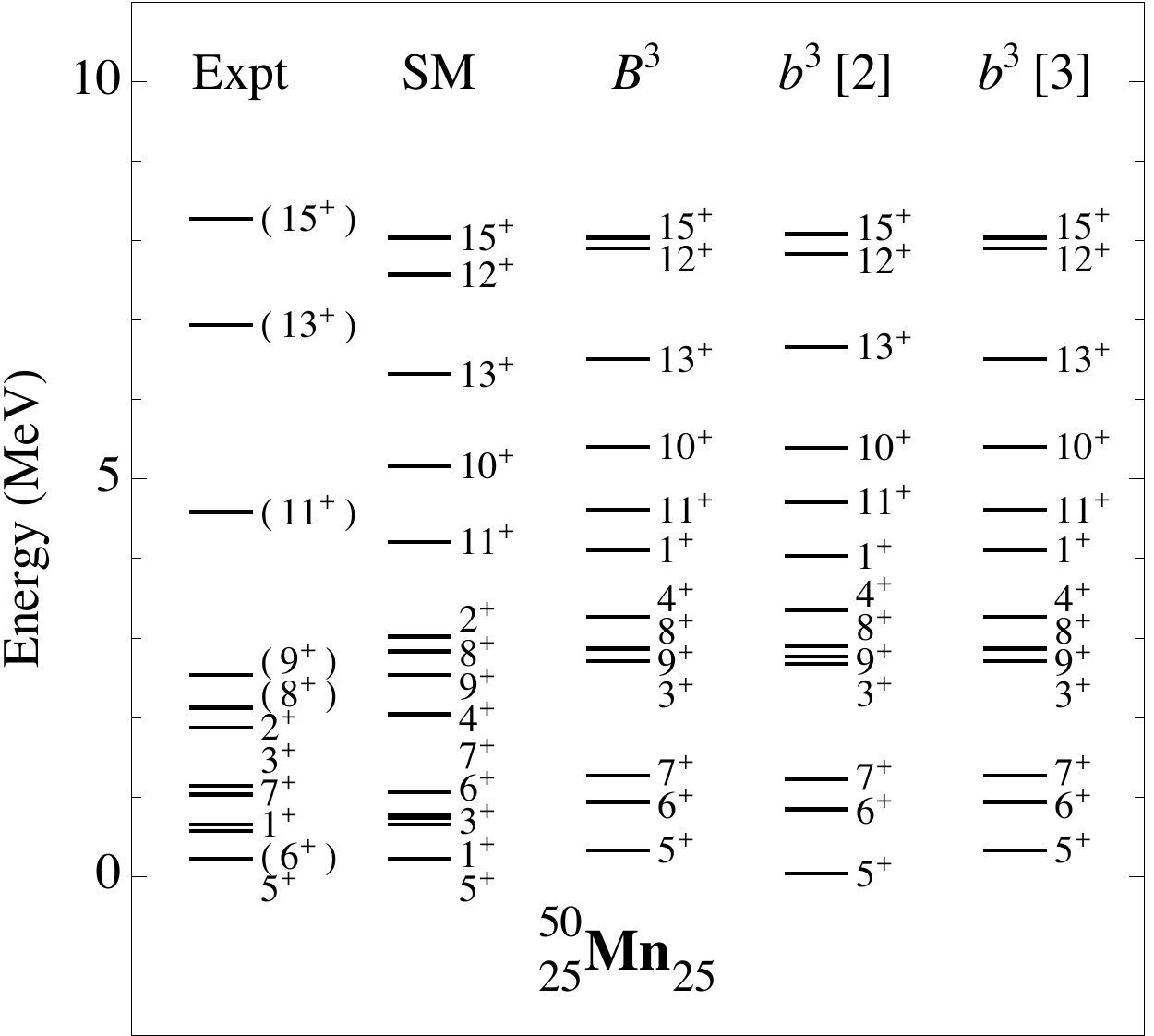}
\caption{
\label{f_spec50}
Same as Fig.~\ref{f_spec46} for $^{50}$Mn.
The shell-model energy of the $T=0$ ground state, $J^\pi=5^+$,
is normalized to the experimental excitation energy of this level
which is relative to the $0^+$ ground state with isospin $T=1$.}
\end{figure}
The results of the wave-function analysis
are confirmed by the energy spectra shown in Figs.~\ref{f_spec46} and~\ref{f_spec50}.
The observed spectra are reasonably well reproduced in the shell-model calculation,
the main deficiency of the latter being that it cannot account for the $3^+$--$5^+$ inversion
of $T=0$ ground states between $^{46}$V and $^{50}$Mn.
The same level of agreement is found in the $B$-pair calculation
except that the low-spin states ($1^+$, $2^+$, $3^+$ and $4^+$)
are at much higher energies (or absent in the case of the $2^+$ level),
in disagreement with the data.

The mapped two-body boson hamiltonian (column `$b^3[2]$')
closely reproduces the $B^3$ calculation,
including its deficient low-spin levels.
Consequently, the three-body components
of the interaction between the $b$ bosons are small.
Let us consider two examples
to illustrate the calculation of three-body interactions between the bosons,
namely those pertaining to the $5^+$ and $7^+$ states.
Numerical values are quoted for $^{46}$V,
the results obtained for $^{50}$Mn being similar.
For $J=5$ there are two independent fermionic $B^3$ states
and the diagonalization of the shell-model hamiltonian in this basis
yields the eigenvalues $\{E_k\}=\{-20.599,-18.456\}$, in MeV.
The same number of independent bosonic $b^3$ states exists,
which can be chosen as $|b^3[\tilde L_2]5\rangle$
with $\tilde L_2=12$ and 2.
The first state in this basis is taken as $\tilde L_2=12$
because its fermionic analogue, $|B^3[\tilde{12}]5\rangle$,
has maximum overlap with the shell-model $5^+_1$ state.
The second state in the boson basis is orthogonal to $|b^3[\tilde{12}]5\rangle$
and therefore unique, and hence its $\tilde L_2$ can be chosen freely.
The diagonalization of the mapped one-plus-two-body boson hamiltonian in this basis
leads to the eigenvalues $\{E'_k\}=\{-20.897,-18.391\}$, in MeV.
The transformation~(\ref{e_bme3b}) of the differences
$\{E_k-E'_k\}=\{0.298,-0.065\}$
back to the orthogonal boson basis
leads to the three-body interaction (in MeV)
\begin{equation}
\langle b^3[\tilde L_2]5|\hat H^{\rm b}_3|b^3[\tilde L'_2]5\rangle=
\left[\begin{array}{rcr}
0.290&&0.053\\
0.053&&-0.057
\end{array}\right],
\qquad
L_2,L'_2=12,2.
\label{e_bme3c}
\end{equation}

The $J=7$ interaction can be dealt with in a similar way.
There are two independent fermionic $B^3$ states
and the diagonalization of the shell-model hamiltonian in the $B$-pair space
leads to the eigenvalues $\{E_k\}=\{-19.639,-16.905\}$, in MeV.
In this case there are {\em three} independent bosonic states $|b^3[\tilde L_2]7\rangle$
and the choice $\tilde L_2=0$, 12 and 2
maximizes the overlap with the shell-model $7^+_1$ state.
The diagonalization of the one-plus-two-body boson hamiltonian in this basis
yields the eigenvalues $\{E'_k\}=\{-19.673,-16.925,+\infty\}$, in MeV.
The spurious state in the three-boson system
is thus removed by the two-body interaction matrix element $\upsilon^{2\rm b}_{14}=+\infty$.
Nevertheless, the entire $3\times3$ matrix must be used
to define the three-body interaction for $J=7$.
This is achieved by transforming the differences
$\{E_k-E'_k\}=\{0.034,-0.020,0.000\}$ back to the orthogonal boson basis,
leading to the three-body interaction (in MeV)
\begin{equation}
\langle b^3[\tilde L_2]7|\hat H^{\rm b}_3|b^3[\tilde L'_2]7\rangle=
\left[\begin{array}{rcrcr}
0.019&~&0.014&&-0.010\\
0.014&&0.020&&0.003\\
-0.010&&0.003&&0.017
\end{array}\right],
\qquad
L_2,L'_2=0,12,2.
\label{e_bme3d}
\end{equation}
Typically, the three-body matrix elements
are of the order of a few tens of keV,
the matrix element $\langle b^3[\tilde{12}]5|\hat H^{\rm b}_3|b^3[\tilde{12}]5\rangle$
in Eq.~(\ref{e_bme3c}) being by far the largest three-body correction in the $1f_{7/2}$ shell.

It will not have escaped the attention of the diligent reader
that the dimensions of all hamiltonian matrices
in the different approximations are small.
The largest dimension,
in the shell-model calculation for six nucleons with $J=3$ or 5 and $T=0$, is twelve.
Modern shell-model codes
usually adopt an $m$-scheme basis
without good angular momentum and isospin
but, even so, dimensions in a single-$j$ shell do remain modest,
of the order of a few hundred at most.
Why then, this diligent reader might well ask, bother
to seek a further reduction of dimension in terms of $B$ pairs
which introduces major computational complications?
The answer is that conceptual insight is gained.

Let us illustrate this with the example of the yrast $5^+$ state
in $^{46}$V or $^{50}$Mn.
According to Fig.~\ref{f_bb_vmn} this state
has a large component in the $B$-pair space
which is of dimension two.
In fact, the analysis of its wave function shows that
$\langle5^+_1|B^3[12]5\rangle^2=0.961~(0.967)$
in $^{46}$V ($^{50}$Mn).
The $5^+_1$ state can therefore be written approximately as $|B^3[12]5\rangle$,
which is nothing but the normalized $B$-pair state~(\ref{e_fnobas})
with $n=3$, $L_2=12$ and $L_3=5$,
and the structure of this state is now understood in simple terms.
For example, within this approximation its energy can be given as
\begin{eqnarray}
E(B^3[12]5)&=&
\frac{7695}{11668}\upsilon^{2\rm f}_{01}+
\frac{564181}{665076}\upsilon^{2\rm f}_{10}+
\frac{6112703}{1551844}\upsilon^{2\rm f}_{21}+
\frac{2544169}{2438612}\upsilon^{2\rm f}_{30}+
\nonumber\\&&
\frac{5814660}{4267571}\upsilon^{2\rm f}_{41}+
\frac{3323}{5834}\upsilon^{2\rm f}_{50}+
\frac{3705457}{1219306}\upsilon^{2\rm f}_{61}+
\frac{340651}{96261}\upsilon^{2\rm f}_{70},
\label{e_ener5a}
\end{eqnarray} 
in terms of the shell-model two-body matrix elements $\upsilon^{2\rm f}_{JT}$.
One notes the large coefficient
in front of the `quadrupole pairing' matrix element $\upsilon^{2\rm f}_{21}$.
Quadrupole collectivity will therefore strongly influence
the energy of the $5^+_1$ level, in both $^{46}$V and $^{50}$Mn.

The derivation of the energy formula~(\ref{e_ener5a}) is non-trivial
since it requires a symbolic implementation of Chen's recursive algorithm~\cite{Chen97},
and overlaps involving up to four pairs are needed [see Eq.~(\ref{e_fme2g})].
On the basis of more `elementary' techniques,
an approximate formula can be obtained as follows.
For a three-boson state, its diagonal energy 
originating from a two-body interaction
can be written with CFPs,
\begin{equation}
V(b^3[L_2]L_3)=
3\sum_{L'_2}
[\ell^2(L'_2)\ell|\}\ell^3[L_2]L_3]^2
\upsilon^{2\rm b}_{L'_2},
\label{e_ener32}
\end{equation}
which are know in closed form in terms of Racah coefficients~\cite{Talmi93},
leading to the expression
\begin{eqnarray}
V(b^3[12]5)&=&
\frac{4370}{11557}\upsilon^{2\rm b}_{2}+
\frac{512325}{392938}\upsilon^{2\rm b}_{4}+
\frac{12650}{41021}\upsilon^{2\rm b}_{6}+
\frac{300}{31369}\upsilon^{2\rm b}_{8}+
\nonumber\\&&
\frac{405}{12265279}\upsilon^{2\rm b}_{10}+
\frac{14859}{14858}\upsilon^{2\rm b}_{12}.
\label{e_ener5b}
\end{eqnarray} 
Since the boson interaction matrix elements $\upsilon^{2\rm b}_L$
are known in terms of the two-body fermion matrix elements $\upsilon^{2\rm f}_{JT}$
from Eq.~(\ref{e_bmej}),
the following total energy (which includes the single-boson energy $3\epsilon_b$)
is found:
\begin{eqnarray}
E(b^3[12]5)&=&3\epsilon_b+V(b^3[12]5)
\nonumber\\&=&
0.891\,\upsilon^{2\rm f}_{01}+
0.888\,\upsilon^{2\rm f}_{10}+
3.637\,\upsilon^{2\rm f}_{21}+
0.981\,\upsilon^{2\rm f}_{30}+
\nonumber\\&&
1.416\,\upsilon^{2\rm f}_{41}+
0.592\,\upsilon^{2\rm f}_{50}+
3.056\,\upsilon^{2\rm f}_{61}+
3.540\,\upsilon^{2\rm f}_{70},
\label{e_ener5c}
\end{eqnarray} 
where the coefficients are rational numbers involving very large integers,
to which a numerical approximation is given.

Equation~(\ref{e_ener5c}) is the boson analogue
of the shell-model result~(\ref{e_ener5a}).
The expressions are similar but not identical,
and this is due to the two-body approximation in the boson calculation.
It should be emphasized once more that,
if three-body interactions between the bosons are included,
results in the $B^3$ and $b^3$ spaces become identical.

The nuclei $^{46}$V and $^{50}$Mn have several isomeric states~\cite{NNDC},
with half-lives ranging from minutes (the $5^+_1$ level in $^{50}$Mn)
to milli- and nano-seconds (the $3^+_1$ and $5^+_1,7^+_1$ levels in $^{46}$V, respectively),
some of which have known dipole and/or quadrupole moments.
The measured magnetic dipole moments
can be compared with the simple single-$j$ shell prediction
that the $g$ factor of any state in an $N=Z$ nucleus equals $(g_\nu+g_\pi)/2$
(see Subsect.~\ref{ss_j}).
The effect of the quenching of the spin part of the M1 operator is small:
without quenching $(g_\nu+g_\pi)/2$ equals 0.55~$\mu_{\rm N}$ in the $1f_{7/2}$ shell
and, with a quenching of 0.7, it reduces to 0.52~$\mu_{\rm N}$.
Therefore, the single-$j$ shell model
predicts magnetic dipole moments $\mu(J^\pi)$
of states in $N=Z$ nuclei in the $1f_{7/2}$ shell
of the order of $0.52J$ to $0.55J$~$\mu_{\rm N}$.
This agrees with the measured~\cite{Sielemann82,Charlwood10} values of
$\mu(3^+_1)=1.64~(3)$~$\mu_{\rm N}$ in $^{46}$V
and $\mu(5^+_1)=2.76~(1)$~$\mu_{\rm N}$ in $^{50}$Mn.
As argued in Subsect.~\ref{ss_j},
this result does not constitute a test of the $B$-pair or $b$-boson approximation,
but shows consistency with a single-$j$ shell truncation.
The large-scale shell-model result with the gxpf1a interaction~\cite{Charlwood10},
$\mu(5^+_1)=2.81$~$\mu_{\rm N}$,
also agrees with the data.

Charlwood {\it et al.}~\cite{Charlwood10}
also measured the quadrupole moment
of the $5^+$ isomer in $^{50}$Mn, $Q(5^+_1)=+0.80~(12)$~b.
In the large-scale shell model with the gxpf1a interaction~\cite{Charlwood10}
one finds a smaller value of $Q(5^+_1)=+0.58$~b.
A numerical calculation in a single-$j$ shell
gives $Q(5^+_1)=+4.2(e_\nu+e_\pi)l_{\rm ho}^2$,
in terms of the neutron (proton) effective charges $e_\nu$ ($e_\pi$)
and the oscillator length $l_{\rm ho}$ of Eq.~(\ref{e_fe2b}).
This shell-model result can be compared
with the approximation in terms of $b$ bosons
that assumes $|5^+_1\rangle\approx|b^3[12]5\rangle$.
One uses the definition
\begin{equation}
Q(b^3[L_2]L_3)=
\sqrt{\frac{16\pi}{5}}
\left(\begin{array}{rcrcr}
L_3&~&2&~&L_3\\-L_3&&0&&L_3
\end{array}\right)
\langle b^3[L_2]L_3||e_b(b^\dag\times\tilde b)^{(2)}||b^3[L_2]L_3\rangle,
\label{e_qm}
\end{equation}
where the reduced matrix element is obtained from
\begin{eqnarray}
\lefteqn{\langle b^3[L_2]L_3||(b^\dag\times\tilde b)^{(\lambda)}||b^3[L'_2]L'_3\rangle=
3(-)^{\ell+L_3+\lambda}\sqrt{(2\lambda+1)(2L_3+1)(2L'_3+1)}}
\nonumber\\&&
\qquad\qquad\times\sum_{L''_2}
[\ell^2(L''_2)\ell|\}\ell^3[L_2]L_3]
[\ell^2(L''_2)\ell|\}\ell^3[L'_2]L'_3]
\left\{\begin{array}{ccccc}
\ell&~&L_3&~&L''_2\\L'_3&&\ell&&\lambda
\end{array}\right\}.
\label{e_qmred}
\end{eqnarray}
For $L_2=12$ and $L_3=5$
and with the effective boson charge taken from Eq.~(\ref{e_eb}),
one finds
\begin{equation}
Q(b^3[12]5)=
-\frac{649485}{150241}(e_\nu+e_\pi)l_{\rm ho}^2\approx
-4.3(e_\nu+e_\pi)l_{\rm ho}^2,
\label{e_q5fin}
\end{equation}
in excellent agreement with the single-$j$ shell result,
considering that a change of sign of the quadrupole moment
is needed to pass from particles to holes~\cite{Lawson80}.

In a single-$j$ shell calculation
the quadrupole deformation is significantly underestimated
if standard values for the effective charges ($e_\nu=0.5$ and $e_\pi=1.5$) are taken.
The dependence of the quadrupole moments
on effective charges and on the oscillator length can be eliminated
by considering ratios.
For example, by making the associations
$|7^+_1\rangle=|b\rangle$ and $|5^+_1\rangle\approx|b^3[12]5\rangle$
for the $7^+_1$ and $5^+_1$ states in $^{42}$Sc and $^{46}$V, respectively,
one obtains the ratio
\begin{equation}
\frac{Q(5^+_1;\,^{46}{\rm V})}{Q(7^+_1;\,^{42}{\rm Sc})}\approx
\frac{Q(b^3[12]5)}{Q(b)}=
\frac{216495}{150241}\approx1.44.
\label{e_q57}
\end{equation}
This is a parameter-independent test
of the validity of the $b$-boson approximation.

\subsubsection{$^{48}$Cr}
\label{sss_cr48}
\begin{figure}
\includegraphics[height=4cm]{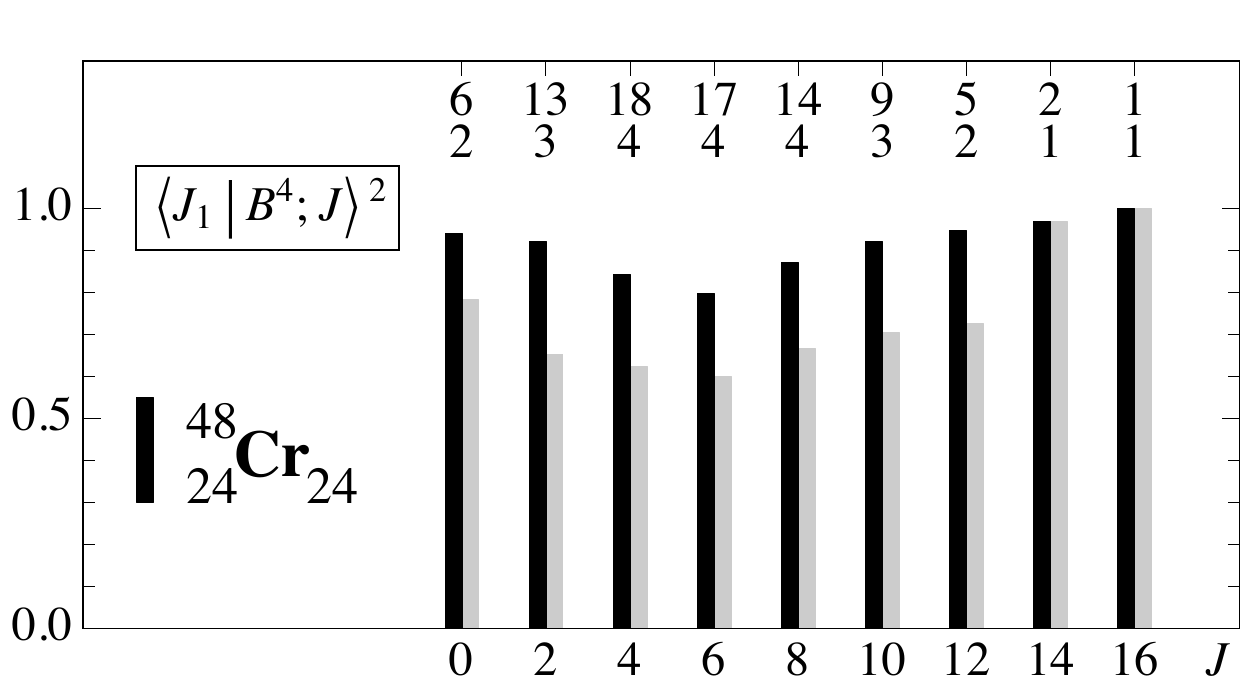}
\caption{
\label{f_bb_cr48}
The square of the projection of the yrast eigenstates in the $(1f_{7/2})^8$ system
onto the subspace spanned by the $B$-pair states $|B^4;J\rangle$,
for angular momentum $J$ and isospin $T=0$.
The shell-model interaction is defined in Eq.~(\ref{e_vf}).
Also shown are the numbers of $(1f_{7/2})^8$ states (top)
and of $B$-pair states (bottom)
with angular momentum $J$ and isospin $T=0$.
The grey bars represent the corresponding analysis
in terms of the stretched configuration of Danos and Gillet~\cite{Danos66,Danos67}
(see text).}
\end{figure}
\begin{figure}
\includegraphics[width=8cm]{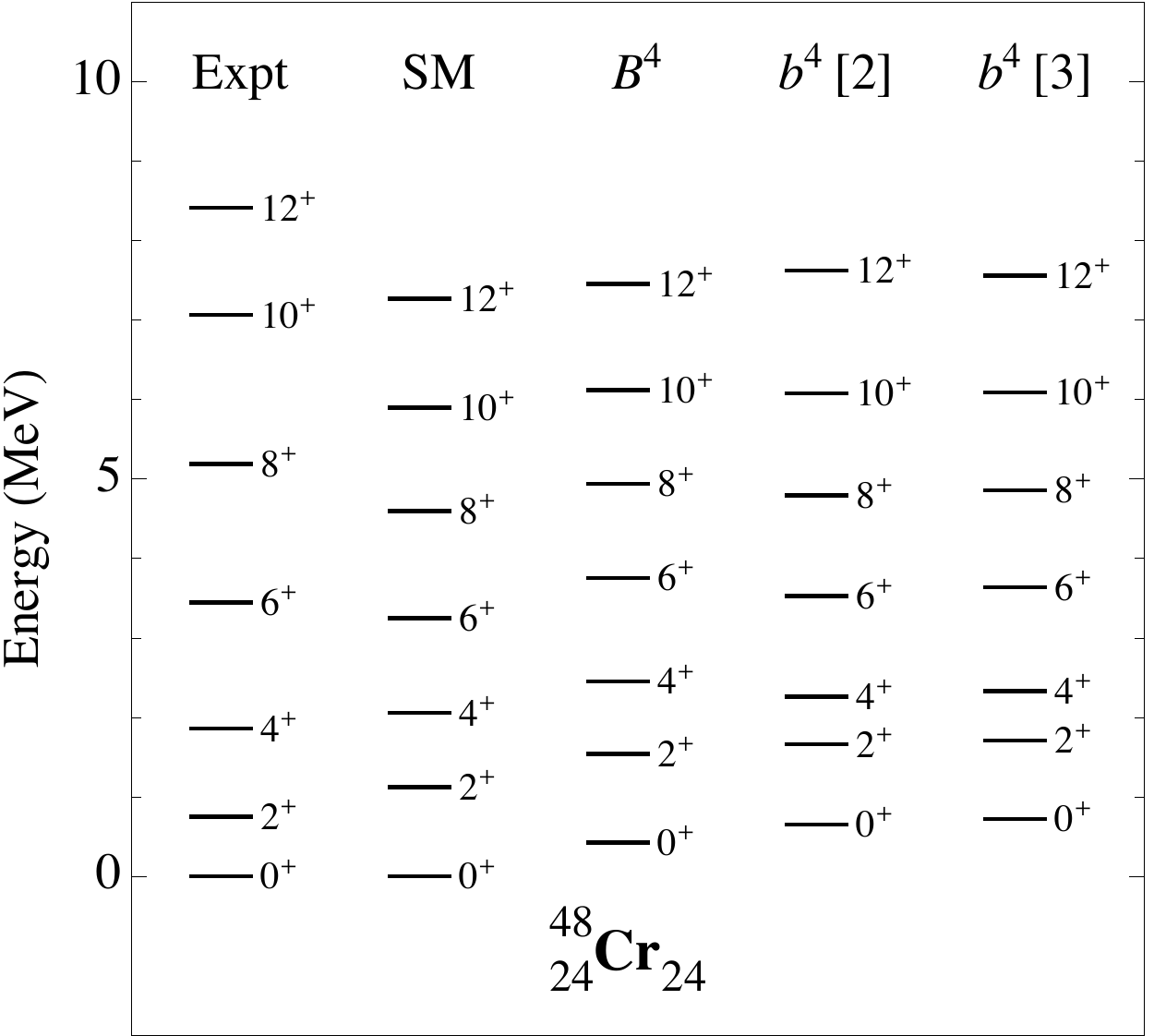}
\caption{
\label{f_spec48}
The yrast spectrum of $^{48}$Cr.
Levels are labelled by their angular momentum and parity $J^\pi$.
The different columns contain the experimental~\cite{NNDC} levels (Expt),
the results of the $(1f_{7/2})^8$ shell model (SM) with the interaction~(\ref{e_vf}),
the lowest eigenvalue of the shell-model hamiltonian in the $B$-pair subspace ($B^4$)
and the lowest eigenvalue of the mapped boson hamiltonian
with up to two-body ($b^3[2]$) and up to three-body ($b^3[3]$) interactions.
The shell-model energy of the $0^+_1$ level is normalized to zero.}
\end{figure}
The $B$-pair content of yrast states in $^{48}$Cr
is displayed in Fig.~\ref{f_bb_cr48}
while its energy spectrum,
calculated in various approximations,
single-$j$ shell model (SM),
shell-model $B$-pair approximation ($B^4$)
and mapped $b$-boson calculation
with up to two-body ($b^4[2]$) and three-body ($b^4[3]$) interactions,
is shown in Fig.~\ref{f_spec48}. 

Two issues of interest arise for the eight-nucleon system.
First, it is possible to establish an explicit connection
with the stretch scheme of Danos and Gillet~\cite{Danos66,Danos67},
since their eight-nucleon stretched state $|B^4_{\rm s}J\rangle$
with angular momentum $J$ 
can in fact be written as
\begin{eqnarray}
|B^4_{\rm s}J\rangle&\propto&
\left(\left(B^\dag\times B^\dag\right)^{(J_{\rm max})}\times
\left(B^\dag\times B^\dag\right)^{(J_{\rm max})}\right)^{(J)}
|{\rm o}\rangle
\nonumber\\&=&(-)^J
\sum_L\sqrt{(2J_{\rm max}+1)(2L+1)}
\left\{\begin{array}{ccccc}
J_{\rm max}&&\ell&&L\\\ell&&J&&J_{\rm max}
\end{array}\right\}
|B^4J_{\rm max}LJ\rangle,
\label{e_stretch}
\end{eqnarray}
in terms of the $B$-pair states~(\ref{e_fnobas}) with $J_{\rm max}=4j-2$.
It is therefore possible to determine the `stretch' content of a shell-model state
since it can be done for the states on the right-hand side of Eq.~(\ref{e_stretch})
with the formalism developed in Sect.~\ref{s_npsm}.
Note that, unlike in the original discussion of Danos and Gillet~\cite{Danos66,Danos67},
anti-symmetry of the stretched configuration is fully taken into account here. 
The stretch content of yrast states in $^{48}$Cr
is shown with grey bars in Fig.~\ref{f_bb_cr48}.
It is clear from Eq.~(\ref{e_stretch}) that the stretched configuration
is but one particular vector in the $B$-pair space
and the stretch content of any state
is therefore necessary smaller than its $B$-pair content.
If the dimension of the $B$-pair space reduces to one,
as is the case for $J=14$ and 16,
both approximations become identical.
These findings are completely at variance with the results of Daley~\cite{Daley87}.

The second question of interest
concerns the formation of a $B$-pair condensate.
The ground state lies dominantly in the $B$-pair space
and can, to a good approximation,
be written in terms of a single component $|B^2[0]B^2[0];0\rangle$
which arises by the coupling of {\em pairs of $B$ pairs} to angular momentum zero.
A wave-function analysis shows that
$\langle0^+_1|B^2[0]B^2[0];0\rangle^2=0.927$,
close to full $B$-pair content of 0.940.
A similar approximation is possible for the $2^+_1$ state
for which $\langle2^+_1|B^2[0]B^2[2];2\rangle^2=0.918$.
In view of these large overlaps, it is then tempting to postulate a seniority-like scheme for the $B$ pairs
[and therefore an SO($2\ell+1$) classification for the $b$ bosons]
but this would be wrong.
Although the $4^+_1$ state has a dominant $B$-pair content (84.2\%),
its $B$-pair seniority-like component is negligible,
$\langle4^+_1|B^2[0]B^2[4];4\rangle^2=0.005$.
The two-body boson interactions,
derived from the shell model and shown in Table~\ref{t_bint},
do not allow an obvious treatment in terms of boson symmetries.
It remains nevertheless true that the single component $|B^2[0]B^2[0];0\rangle$
provides a good approximation to the {\em ground state}
of the eight-nucleon system.
It would be of some interest to generalize this finding
to larger single-$j$ shells and to many particles.

\subsection{The $1g_{9/2}$ shell}
\label{ss_g9}
Nuclei in the $1g_{9/2}$ shell were the focus of a previous study~\cite{Zerguine11},
with a wave-function analysis limited to $n=4$
and boson interactions limited to two-body.
Additional material and further details of the calculations
are presented in the subsequent subsections.
The shell-model interaction, referred to as SLGT0, is taken 
from Serduke {\it et al.}~\cite{Serduke76}
and gives satisfactory results
for the neutron-deficient nuclei in the mass region $A=86$ to 100~\cite{Herndl97}.
Being defined in the $2p_{1/2}+1g_{9/2}$ shell-model space,
this interaction is renormalized to the $1g_{9/2}$ orbit.
The resulting matrix elements are given in Table~\ref{t_fint}.

\subsubsection{$^{96}$Cd}
\label{sss_cd96}
\begin{figure}
\includegraphics[height=4cm]{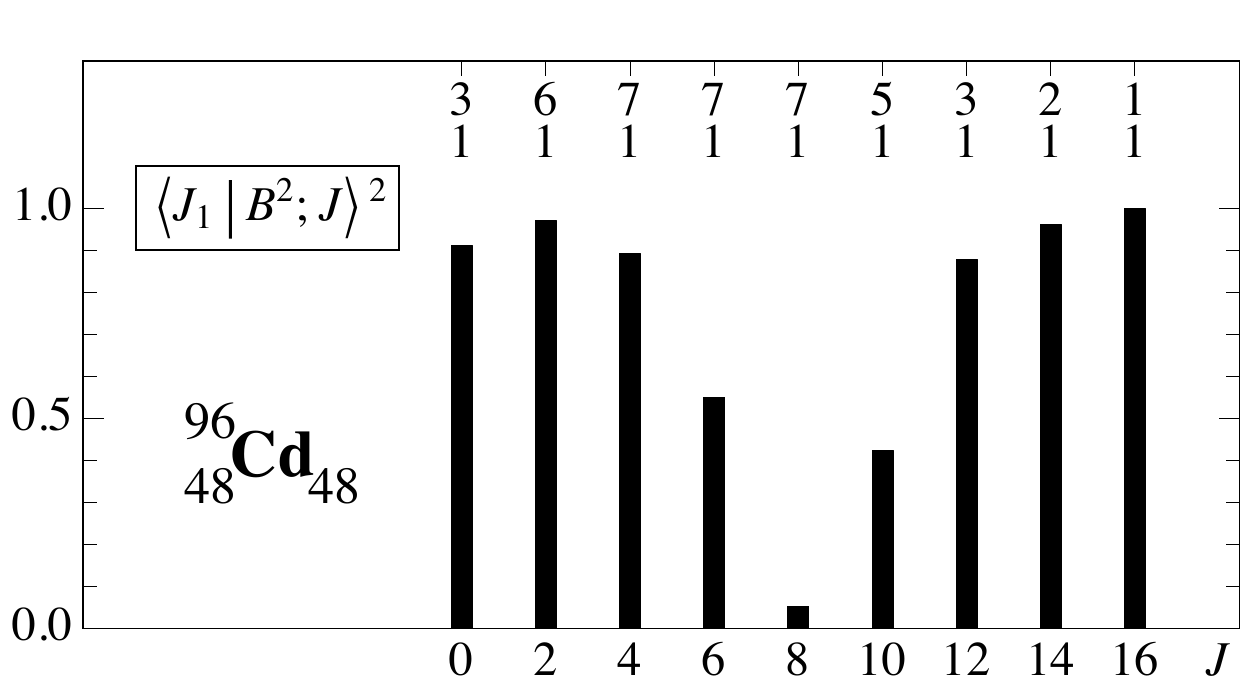}
\caption{
\label{f_bb_cd96}
Overlaps of the yrast eigenstates in the $(1g_{9/2})^4$ system,
for angular momentum $J$ and isospin $T=0$,
with the $B$-pair state $|B^2;J\rangle$.
The shell-model interaction SLGT0 is defined in Table~\ref{t_fint}.
Also shown are the numbers of $(1g_{9/2})^4$ states (top)
and of $B$-pair states (bottom)
with angular momentum $J$ and isospin $T=0$.}
\end{figure}
The study of this nucleus is at the limit of present experimental capabilities.
A fusion--evaporation experiment was proposed at GANIL some time ago~\cite{Cederwall11un}
but had to be rescheduled to due to technical difficulties.
In view of this current interest,
it is worthwhile to investigate the $B$-pair structure of $^{96}$Cd.
The $B$-pair content of shell-model states calculated with the SLGT0 interaction
is shown in Fig.~\ref{f_bb_cd96}.
Results are entirely consistent with those
obtained in the $1f_{7/2}$ shell (see Fig.~\ref{f_bb_tife}),
indicating the generic nature of the analysis,
independent of the particular value of $j$ of the shell considered.
The decrease of the $B$-pair content
at intermediate values of the angular momentum $J$
can be understood on the basis of a combination of geometry---the
CFPs in a single-$j$ shell, and dynamics---the
dependence of the interaction matrix elements on $J$ and $T$~\cite{Zerguine11}.

\begin{figure}
\includegraphics[width=8cm]{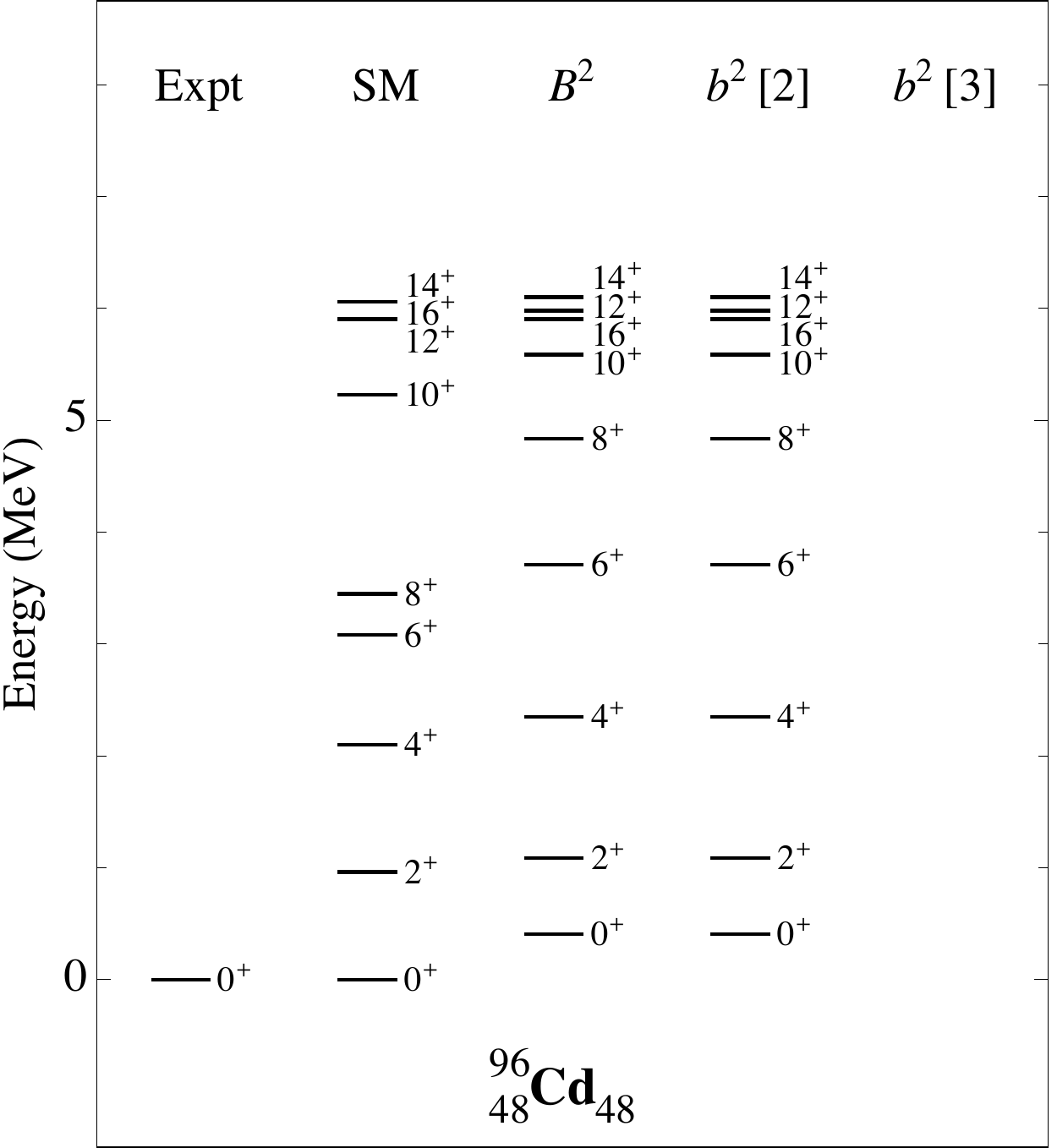}
\caption{
\label{f_spec96}
The yrast spectrum of $^{96}$Cd.
Levels are labelled by their angular momentum and parity $J^\pi$.
No experimental levels are known to date.
The different columns contain
the results of the $(1g_{9/2})^4$ shell model (SM) with the SLGT0 interaction,
the expectation value of the shell-model hamiltonian in the $B$-pair state ($B^2$)
and the expectation value of the mapped boson hamiltonian
with up to two-body interactions ($b^2[2]$).
The shell-model energy of the $0^+_1$ level is normalized to zero.}
\end{figure}
\begin{table}[pt]
\tbl{\label{t_bmeg}
Coefficients $a^L_{JT}(j)$ in the expansion~(\ref{e_bmej}) for $j=9/2$.}
{\begin{tabular}{@{}ccccccccccc@{}}
\toprule
&\hspace{-15pt}$L$\hspace{-15pt}&0&2&4&6&8&10&12&14&16\\[-2pt]
$(JT)$&&&&&&&&&&\\[-2pt]
\colrule
$(01)$&&
$\frac{4862}{4265}$&
$\phantom{-}\frac{117572}{137755}$&
$\frac{2261}{5445}$&
$\phantom{-}\frac{7429}{65120}$&
$\frac{1311}{110210}$&&&&\\[3pt]
$(10)$&&
$\frac{35802}{46915}$&
$\phantom{-}\frac{976752}{1515305}$&
$\frac{2261}{5445}$&
$\phantom{-}\frac{11799}{65120}$&
$\frac{50301}{1212310}$&
$\frac{345}{156739}$&&&\\[3pt]
$(21)$&&
$\frac{15912}{9383}$&
$\phantom{-}\frac{543932}{303061}$&
$\frac{13889}{7986}$&
$\phantom{-}\frac{45011}{35816}$&
$\frac{67068}{121231}$&
$\frac{15525}{156739}$&&&\\[3pt]
$(30)$&&
$\frac{139944}{609895}$&
$\phantom{-}\frac{6520724}{19698965}$&
$\frac{771799}{1557270}$&
$\phantom{-}\frac{165669}{291005}$&
$\frac{3425436}{7880015}$&
$\frac{359415}{2037607}$&
$\frac{1218}{69355}$&&\\[3pt]
$(41)$&&
$\frac{99144}{609895}$&
$\phantom{-}\frac{6750054}{19698965}$&
$\frac{618032}{778635}$&
$\phantom{-}\frac{821583}{582010}$&
$\frac{13883076}{7880015}$&
$\frac{2774250}{2037607}$&
$\frac{63423}{138710}$&&\\[3pt]
$(50)$&&
$\frac{408}{55445}$&
$\phantom{-}\frac{43358}{1790815}$&
$\frac{686}{7865}$&
$\phantom{-}\frac{483}{2035}$&
$\frac{334476}{716365}$&
$\frac{578322}{926185}$&
$\frac{29957}{63050}$&
$\frac{868}{8515}$&\\[3pt]
$(61)$&&
$\frac{153}{93830}$&
$\phantom{-}\frac{13699}{1515305}$&
$\frac{343}{6655}$&
$\phantom{-}\frac{76797}{358160}$&
$\frac{789021}{1212310}$&
$\frac{1113966}{783695}$&
$\frac{109881}{53350}$&
$\frac{1953}{1310}$&\\[3pt]
$(70)$&&
$\frac{81}{4147286}$&
$\phantom{-}\frac{13608}{66976481}$&
$\frac{1673}{882453}$&
$\phantom{-}\frac{14823}{1217744}$&
$\frac{3024621}{53584102}$&
$\frac{6727847}{34639319}$&
$\frac{1148337}{2358070}$&
$\frac{46251}{57902}$&
$\frac{8}{17}$\\[3pt]
$(81)$&&
$\frac{1}{1219790}$&
$\phantom{-}\frac{378}{19698965}$&
$\frac{161}{519090}$&
$\phantom{-}\frac{29223}{9312160}$&
$\frac{175176}{7880015}$&
$\frac{1202172}{10188035}$&
$\frac{15231}{31525}$&
$\frac{1977}{1310}$&3\\[3pt]
$(90)$&&
$\frac{1}{24310}$&
$-\frac{8081}{334882405}$&
$\frac{23}{802230}$&
$-\frac{101}{14391520}$&
$\frac{61018}{133960255}$&
$\frac{479646}{173196595}$&
$\frac{10893}{535925}$&
$\frac{2211}{22270}$&
$\frac{9}{17}$\\[3pt]
\botrule
\end{tabular}}
\end{table}
The energy spectrum of $^{96}$Cd,
calculated in various approximations,
single-$j$ shell model (SM),
shell-model $B$-pair approximation ($B^2$)
and mapped $b$-boson calculation ($b^2[2]$),
is shown in Fig.~\ref{f_spec96}.
Results are seen to be consistent with the wave-function analysis.
The boson interaction matrix elements $\upsilon^{2\rm b}_L$
are known analytically
in terms of the two-body fermion matrix elements $\upsilon^{2\rm f}_{JT}$
[see Eq.~(\ref{e_bmej})]
with universal coefficients $a^L_{JT}(j=9/2)$ given in Table~\ref{t_bmeg}.
The resulting boson interaction matrix elements
are shown in Table~\ref{t_bint}.
There is no four-particle shell-model state with $J=18$,
implying the choice $\upsilon^{2\rm b}_{18}=+\infty$.

\subsubsection{$^{94}$Ag}
\label{sss_ag94}
Not much is known experimentally about $^{94}$Ag 
except for the presence of two isomers,
with tentative assignments $J^\pi=7^+$ (presumably the lowest $T=0$ state)
and $J^\pi=21^+$,
the latter at $6.7~(5)$~MeV above the $0^+$ ground state~\cite{Mukha05,Cerny09}.
The subsequent discussion is focussed
on the structure of these two states.

\begin{figure}
\includegraphics[height=4cm]{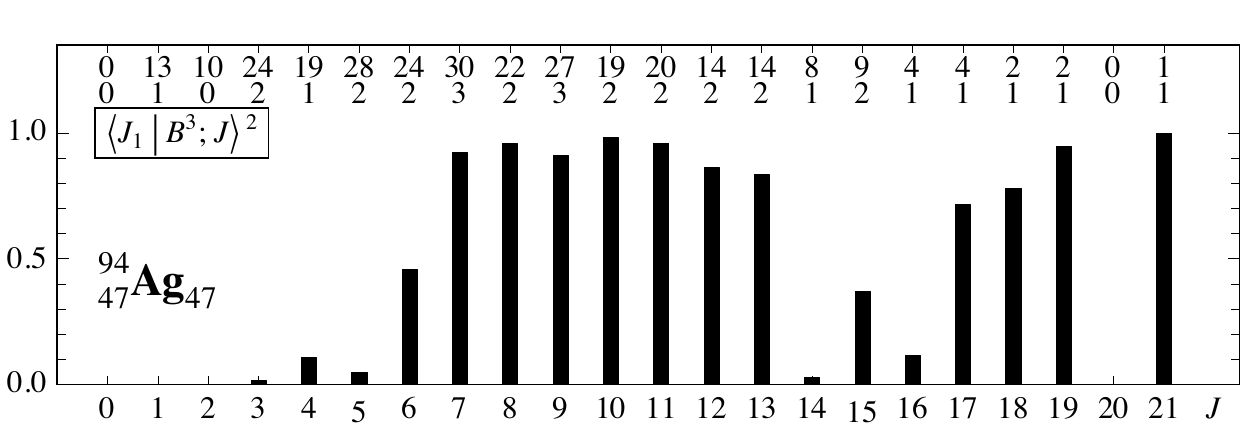}
\caption{
\label{f_bb_ag94}
The square of the projection of the yrast eigenstates in the $(1g_{9/2})^6$ system
onto the subspace spanned by the $B$-pair states $|B^3;J\rangle$,
for angular momentum $J$ and isospin $T=0$.
The shell-model interaction SLGT0 is defined in Table~\ref{t_fint}.
Also shown are the numbers of $(1g_{9/2})^6$ states (top)
and of $B$-pair states (bottom)
with angular momentum $J$ and isospin $T=0$.}
\end{figure}
The $B$-pair content is obtained from Eq.~(\ref{e_fana}) with $\omega$ up to 3,
the maximum dimension of the $B$-pair space (for $J=7$ and 9).
This quantity is shown in Fig.~\ref{f_bb_ag94} for yrast states in $^{94}$Ag,
together with the dimensions of the shell-model and $B$-pair spaces.
The results are in total accord with those found in the $1f_{7/2}$ shell
(see Fig.~\ref{f_bb_vmn}),
considering that the replacement $j=7/2\rightarrow9/2$
leads to an overall increase of the angular momenta involved.
It is seen in particular that the overlaps are high for $J=21$ (which is trivial)
and for $J=7$ (which is not),
making the analysis of these states in terms of $B$ pairs or $b$ bosons meaningful.

\begin{figure}
\includegraphics[width=8cm]{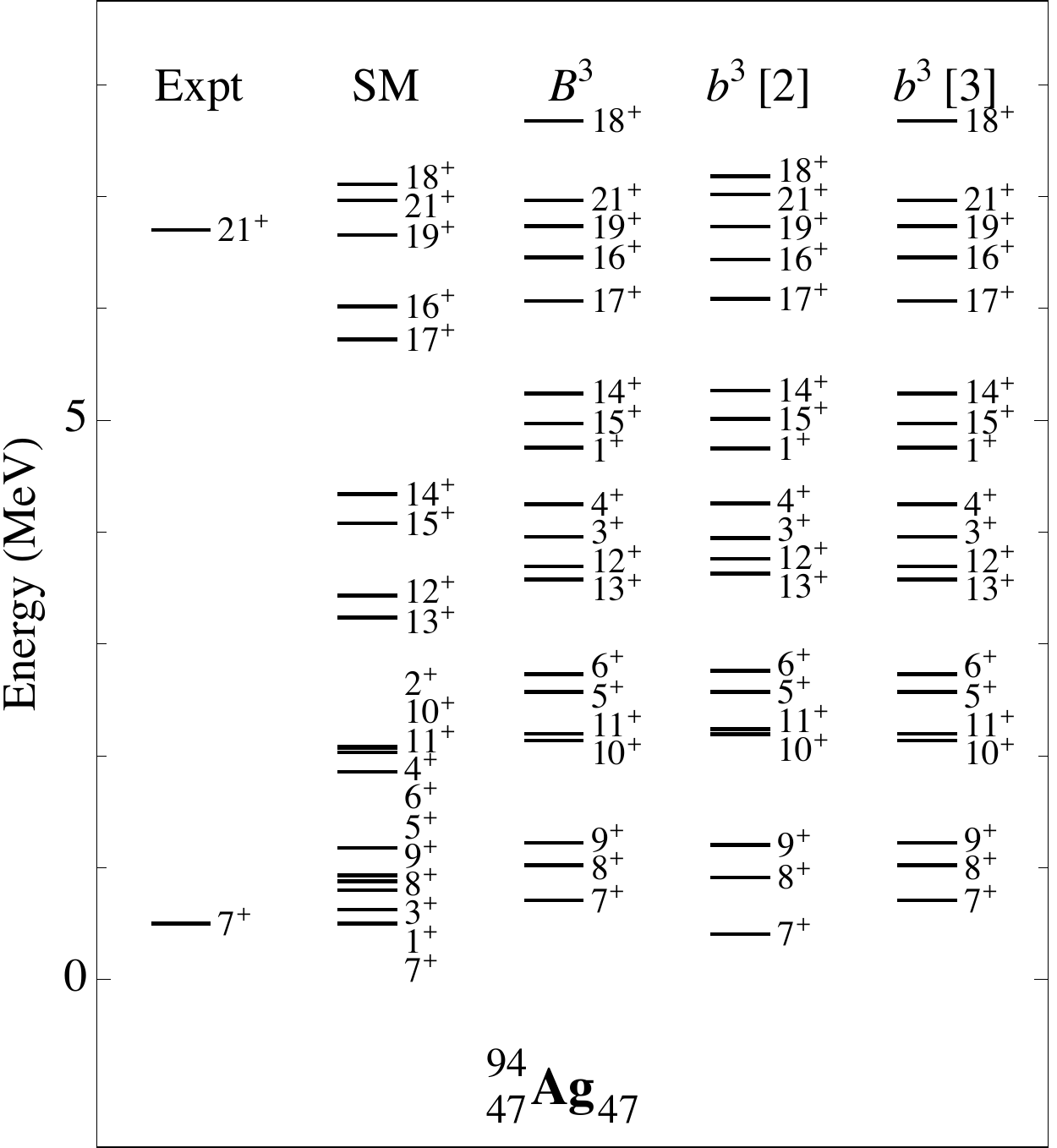}
\caption{
\label{f_spec94}
The spectrum of yrast states in $^{94}$Ag with isospin $T=0$.
Levels are labelled by their angular momentum and parity $J^\pi$.
The different columns contain the experimental~\cite{Mukha05,Cerny09} levels (Expt),
the results of the $(1g_{9/2})^6$ shell model (SM) with the SLGT0 interaction,
the lowest eigenvalue of the shell-model hamiltonian in the $B$-pair subspace ($B^3$)
and the lowest eigenvalue of the mapped boson hamiltonian
with up to two-body ($b^3[2]$) and up to three-body ($b^3[3]$) interactions.
The shell-model energy of the $T=0$ ground state, $J^\pi=7^+$,
is normalized to the experimental excitation energy of this level
which is relative to the $0^+$ ground state.}
\end{figure}
The energy spectrum of $T=0$ states in $^{94}$Ag is shown in Fig.~\ref{f_spec94}.
The shell model with the SLGT0 interaction
gives the correct $T=0$ ground-state spin, $J=7$,
and the energy of the $21^+$ isomer
comes out reasonably close to the its observed value.
The $B$-pair calculation agrees with the shell model
but for the low-spin states ($1^+$ to $5^+$)
which are obtained at much higher energies.

In a shell-model description
where three neutrons and three protons are placed in the $1g_{9/2}$ orbit,
the $J=21$ state is stretched and therefore unique.
In this single-$j$ shell approximation,
the $21^+$ isomer can therefore be written exactly as $|B^3[12]21\rangle$,
the normalized $B$-pair state~(\ref{e_fnobas})
with $n=3$, $L_2=12$ and $L_3=21$.
Chen's algorithm~\cite{Chen97} then provides
the following energy expression for this state:
\begin{equation}
E(B^3[12]21)=
\frac{21}{65}\upsilon^{2\rm f}_{50}+
\frac{21}{10}\upsilon^{2\rm f}_{61}+
\frac{645}{442}\upsilon^{2\rm f}_{70}+
\frac{69}{10}\upsilon^{2\rm f}_{81}+
\frac{717}{170}\upsilon^{2\rm f}_{90},
\label{e_ener21}
\end{equation} 
in terms of the two-body fermion matrix elements $\upsilon^{2\rm f}_{JT}$.
This energy expression can alternatively (and more simply)
be derived with standard techniques involving CFPs.
Since a six-nucleon state
with angular momentum $J_6=21$ and isospin $T_6=0$ is unique,
its energy is given as
\begin{equation}
E(j^6J_6=21,T_6=0)=\sum_{JT} a_{JT}\upsilon^{2\rm f}_{JT},
\label{e_ener21b}
\end{equation}
with the coefficients $a_{JT}$
known in terms of $6\rightarrow4$ CFPs~\cite{Talmi93},
\begin{equation}
a_{JT}=15
\sum_{\alpha_4J_4T_4}
[j^4(\alpha_4J_4T_4)j^2(JT)|\}j^6J_6=21,T_6=0]^2.
\label{e_ener21c}
\end{equation}
It can be verified that this alternative derivation
also yields the expression~(\ref{e_ener21}),
which provides a rigorous check on the correctness
of the implementation of Chen's algorithm.
It should be emphasized that the derivation using standard CFP techniques
is valid only for shell-model states that are unique,
such as the $21^+$ isomer.
If several states can be constructed for a given $J$ and $T$,
no such derivation is possible
while an expression still can be found from $B$ pairs,
as illustrated below for the $7^+$ isomer.

In terms of bosons, the $21^+$ isomer arises
from the coupling of three $b$ bosons with spin $\ell=9$
to total angular momentum $J=21$.
It can be shown~\cite{Isacker12}
that {\em two} independent boson states exist with $J=21$,
one of which must be spurious.
Let us consider this case in detail,
to illustrate the mechanism by which spurious states
can be eliminated analytically.
The two independent boson states may be chosen as
$|b^3[\tilde{12}]21\rangle$ and $|b^3[\tilde{14}]21\rangle$,
assumed to be normalized and orthogonal.
Since the CFPs needed in a three-particle problem
are known in terms of Racah coefficients~\cite{Talmi93},
the energy matrix can be shown to have the following elements:
\begin{eqnarray}
\langle b^3[\tilde{12}]21|\hat H^{\rm b}_2|b^3[\tilde{12}]21\rangle&=&
3\epsilon_b+
\frac{2833}{2697}\upsilon^{2\rm b}_{12}+
\frac{44200}{263469}\upsilon^{2\rm b}_{14}+
\frac{2782494}{2546867}\upsilon^{2\rm b}_{16}+
\frac{60536}{87823}\upsilon^{2\rm b}_{18},
\nonumber\\
\langle b^3[\tilde{14}]21|\hat H^{\rm b}_2|b^3[\tilde{14}]21\rangle&=&
3\epsilon_b+
\frac{3337047}{1932106}\upsilon^{2\rm b}_{14}+
\frac{82824}{87823}\upsilon^{2\rm b}_{16}+
\frac{20553}{62326}\upsilon^{2\rm b}_{18},
\label{e_ener21d}\\
\langle b^3[\tilde{12}]21|\hat H^{\rm b}_2|b^3[\tilde{14}]21\rangle&=&
\sqrt\frac{24582912900}{84841672619}\upsilon^{2\rm b}_{14}-
\sqrt\frac{7946802864}{7712879329}\upsilon^{2\rm b}_{16}+
\sqrt\frac{20067684 }{88284779}\upsilon^{2\rm b}_{18},
\nonumber
\end{eqnarray}
where $\epsilon_b$ is the energy of the $b$ boson.
The stretched boson interaction matrix element $\upsilon^{2\rm b}_{18}$
appears in the diagonal and the off-diagonal matrix elements
and, consequently, in the limit $\upsilon^{2\rm b}_{18}\rightarrow+\infty$,
one eigenvalue of the $2\times2$ matrix~(\ref{e_ener21d}) tends to infinity
while the lowest eigenvalue acquires the expression
\begin{equation}
E(b^3_\infty21)=3\epsilon_b+
\frac{6851}{20155}\upsilon^{2\rm b}_{12}+
\frac{15488}{21545}\upsilon^{2\rm b}_{14}+
\frac{1212882}{624805}\upsilon^{2\rm b}_{16},
\label{e_ener21e}
\end{equation}
where the index `$\infty$' serves
as a reminder of the limit procedure used to derive the result.
The procedure also yields the components of the state,
\begin{equation}
|b^3_\infty21\rangle=
-\sqrt{\frac{637143}{1968935}}|b^3[\tilde{12}]21\rangle+
\sqrt{\frac{1331792}{1968935}}|b^3[\tilde{14}]21\rangle,
\label{e_wav21}
\end{equation}
of use in the calculation of the quadrupole moment of the $21^+$ isomer (see below).
Since the $b$-boson energy and the boson interaction
are known in terms of the two-body fermion interaction,
Eq.~(\ref{e_ener21e}) can be converted into
\begin{eqnarray}
E(b^3_\infty21)&=&
\frac{22134}{3707825}\upsilon^{2\rm f}_{30}+
\frac{1152549}{7415650}\upsilon^{2\rm f}_{41}+
\frac{1347751953}{5740387250}\upsilon^{2\rm f}_{50}+
\frac{8606149749}{4857250750}\upsilon^{2\rm f}_{61}+
\nonumber\\&&
\frac{354940047213}{214690483150}\upsilon^{2\rm f}_{70}+
\frac{1561553973}{220784125}\upsilon^{2\rm f}_{81}+
\frac{15411107094}{3753330125}\upsilon^{2\rm f}_{90}.
\label{e_ener21f}
\end{eqnarray}
This is an approximate expression
since it is derived by use of a mapping
that includes up to two-body interactions between the bosons.
To what extent it is wrong
therefore yields an idea about the reliability of the two-body boson mapping.
Since the highest allowed angular momentum
for two neutrons and two protons in a $j=9/2$ orbit is $J=16$,
only matrix elements $\upsilon^{2\rm f}_{JT}$ with $J\geq5$
can contribute to the energy of the $J=21$ state.
This rule is obviously obeyed in Eq.~(\ref{e_ener21}) but violated in Eq.~(\ref{e_ener21f}).
It is seen, however, that the coefficients of $\upsilon^{2\rm f}_{30}$ and $\upsilon^{2\rm f}_{41}$
are rather small in the latter expression,
indicating that the two-body boson approximation is reasonably accurate.

The perplexed reader might well wonder
what could be the purpose of quoting in Eq.~(\ref{e_ener21f})
the coefficients $a_{JT}$ as the ratio of two ridiculously large integers.
The advantage of the use of {\em exact} numbers
is that enables a rigorous check
of fermionic as well as bosonic calculations.
The coefficients $a_{JT}$ in an energy expression
$E(j^nJ_nT_n)=\sum_{JT} a_{JT}\upsilon^{2\rm f}_{JT}$
for a unique $n$-particle shell-model state $j^n$
with total angular momentum $J_n$ and isospin $T_n$,
satisfy the identities
\begin{eqnarray}
\sum_{JT}a_{JT}&=&
\frac{n(n-1)}{2},
\nonumber\\
\sum_{JT}J(J+1)a_{JT}&=&
J_n(J_n+1)+j(j+1)\times n(n-2),
\nonumber\\
\sum_{JT}T(T+1)a_{JT}&=&
T_n(T_n+1)+{\frac34}n(n-2).
\end{eqnarray}
These identities reflect the conservation
of particle number, angular momentum and isospin
in the shell model
and are therefore valid for the coefficients in Eq.~(\ref{e_ener21}).
They are also {\em exactly} satisfied by the coefficients in Eq.~(\ref{e_ener21f}).
This is a consequence of the preservation of $n$, $J$ and $T$ under the mapping procedure.

The yrast $7^+$ state in $^{94}$Ag
is the analogue of the $5^+$ state in $^{46}$V or $^{50}$Mn,
discussed in Subsect.~\ref{sss_vmn}.
Its structure is particularly simple
since a wave-function analysis shows that
$\langle7^+_1|B^3[16]7\rangle^2=0.908$.
The $7^+$ isomer is now understood in simple terms
as it results from the coupling of two $B$ pairs
to maximal angular momentum $J=16$
($J=18$ is not allowed by the Pauli principle)
which is subsequently coupled with the third $B$ pair to total $J=7$.
Within this approximation its energy is calculated as
\begin{eqnarray}
E(B^3[16]7)&=&
0.528\upsilon^{2\rm f}_{01}+
0.654\upsilon^{2\rm f}_{10}+
3.302\upsilon^{2\rm f}_{21}+
1.081\upsilon^{2\rm f}_{30}+
1.977\upsilon^{2\rm f}_{41}+
\nonumber\\&&
0.256\upsilon^{2\rm f}_{50}+
0.190\upsilon^{2\rm f}_{61}+
0.480\upsilon^{2\rm f}_{70}+
3.002\upsilon^{2\rm f}_{81}+
3.529\upsilon^{2\rm f}_{90}.
\label{e_ener7a}
\end{eqnarray}
For the SLGT0 interaction
this formula gives an energy of $-11.069$~MeV,
to be compared with a correlation energy of $-11.276$~MeV
if the full $(1g_{9/2})^6$ shell-model basis is used.
Note also that the approximate energy expression
for the $7^+_1$ state in the $(1g_{9/2})^6$ system
is similar to the one obtained in Eq.~(\ref{e_ener5a})
for the $5^+_1$ state in the $(1f_{7/2})^6$ system.

Since the dimension of the $B$-pair space is three
and equals the number of independent states for three $b$ bosons with spin $\ell=9$,
no spurious boson states occur for $J=7$.
The calculation of the energy of the three-boson state $|b^3[16]7\rangle$
in a two-body boson approximation is then straightforward
and proceeds along the lines of the energy calculation
for the $5^+$ state in Subsect.~\ref{sss_vmn}
[see Eqs.~(\ref{e_ener32}), (\ref{e_ener5b}) and (\ref{e_ener5c})],
leading to the expression
\begin{eqnarray}
E(b^3[16]7)&=&
0.711\upsilon^{2\rm f}_{01}+
0.711\upsilon^{2\rm f}_{10}+
3.118\upsilon^{2\rm f}_{21}+
0.997\upsilon^{2\rm f}_{30}+
1.933\upsilon^{2\rm f}_{41}+
\nonumber\\&&
0.278\upsilon^{2\rm f}_{50}+
0.235\upsilon^{2\rm f}_{61}+
0.484\upsilon^{2\rm f}_{70}+
3.004\upsilon^{2\rm f}_{81}+
3.529\upsilon^{2\rm f}_{90},
\label{e_ener7b}
\end{eqnarray}
in close correspondence with the fermion result~(\ref{e_ener7a})
that takes into account the exchange terms between the $B$ pairs.

The discussion of Subsect.~\ref{ss_j} concerning magnetic dipole moments
also applies to the $1g_{9/2}$ shell.
The single-$j$ shell prediction for the $g$ factor
of any state in a $1g_{9/2}$ $N=Z$ nucleus
is 0.54~$\mu_{\rm N}$ without spin quenching
and 0.51~$\mu_{\rm N}$ with a quenching of 0.7.
Magnetic dipole moments do not provide a test
of the $B$-pair or $b$-boson approximation
but measured $\mu$ values that deviate from the narrow range
predicted in a single-$j$ shell,
would be indicative
of admixtures of configurations beyond the $1g_{9/2}$ shell.

It is also of interest to predict the quadrupole moments of the isomeric states in $^{94}$Ag.
The shell-model calculation in a single-$j$ approximation
can be worked analytically for $J=21$
since the state is unique,
\begin{equation}
Q(B^3[12]21)=
-\sqrt{\frac{196}{6}}(e_\nu+e_\pi)l_{\rm ho}^2\approx
-0.42~{\rm b}.
\label{e_q21sm}
\end{equation}
The corresponding boson result is obtained from the expansion~(\ref{e_wav21}),
together with the general expressions~(\ref{e_qm}) and~(\ref{e_qmred}),
leading to
\begin{equation}
Q(b^3_\infty21)=
-\sqrt{\frac{81949367824}{3489855625}}(e_\nu+e_\pi)l_{\rm ho}^2\approx
-0.44~{\rm b},
\label{e_q21b}
\end{equation}
which illustrates the reliability of the boson approximation.
The shell-model value of the quadrupole moment of the $7^+$ isomer
can be obtained numerically and, in a single-$j$ shell approximation,
gives $Q(7^+_1)=+6.60(e_\nu+e_\pi)l_{\rm ho}^2\approx+0.60$~b.
The dominant component of this state in terms of $b$ bosons
is $|b^3[16]7\rangle$ for which, from Eqs.~(\ref{e_qm}) and~(\ref{e_qmred}),
the following quadrupole moment is found:
\begin{equation}
Q(b^3[16]7)=
-\sqrt{\frac{30930277300923364}{627253477610841}}(e_\nu+e_\pi)l_{\rm ho}^2\approx
-0.64~{\rm b}.
\label{e_q7b}
\end{equation}
The quadrupole moments~(\ref{e_q21sm}), (\ref{e_q21b}) and~(\ref{e_q7b})
are given for particle--particle configurations;
an additional sign is needed to pass to the hole--hole nucleus $^{94}$Ag.

A final word is needed on the nature of boson approximation.
Consider as an example the matrix element
\begin{equation}
\langle B^3[4]7|B^3[16]7\rangle=
\sqrt{\frac{112919600563049280}{139849953265085321}}\approx0.899,
\label{e_lapB}
\end{equation}
where it is assumed that bra and ket states
are normalized but evidently non-orthogonal.
As can be expected from a fraction
which involves very large integers,
the calculation of this overlap is non-trivial.
The corresponding boson result,
\begin{equation}
\langle b^3[4]7|b^3[16]7\rangle=
\sqrt{\frac{7012200}{8733503}}\approx0.896,
\label{e_lapb}
\end{equation}
is obtained much more simply in terms of $3\rightarrow2$ CFPs
associated with bosons with spin $\ell=9$.
The reliability of the mapping of $B$ pairs onto $b$ bosons
ultimately is due to the negligible effect of exchange terms between the $B$ pairs.
It cannot be emphasized enough
that the calculation of matrix elements of the type~(\ref{e_lapB})
is highly non-trivial and quickly runs into computational problems
as the number of pairs increases.
By comparison, the calculation of overlaps of the type~(\ref{e_lapb})
is trivial and can easily be done for all cases of relevance.

\subsubsection{$^{92}$Pd}
\label{sss_pd92}
The low-lying yrast states of this nucleus
were measured by Cederwall {\it et al.}~\cite{Cederwall11}
with the aim to probe the importance of aligned neutron--proton pairs
in $N\sim Z$ nuclei.
An analysis of shell-model wave functions in terms of $B$ pairs,
as performed for all $N=Z$ nuclei previously considered,
becomes tedious in this case,
owing to the dimensions of complete bases $|\bar P^n_r\rangle$
in terms of pairs $P^\dag_{\Gamma M_\Gamma}$.
The complete study of the $1f_{7/2}$ shell presented in Subsect.~\ref{ss_f7}
and the results obtained so far for the $1g_{9/2}$ shell
indicate that an analysis of $^{92}$Pd in the $B$-pair subspace
and a subsequent mapping to $b$ bosons
should give meaningful results.

\begin{figure}
\includegraphics[width=8cm]{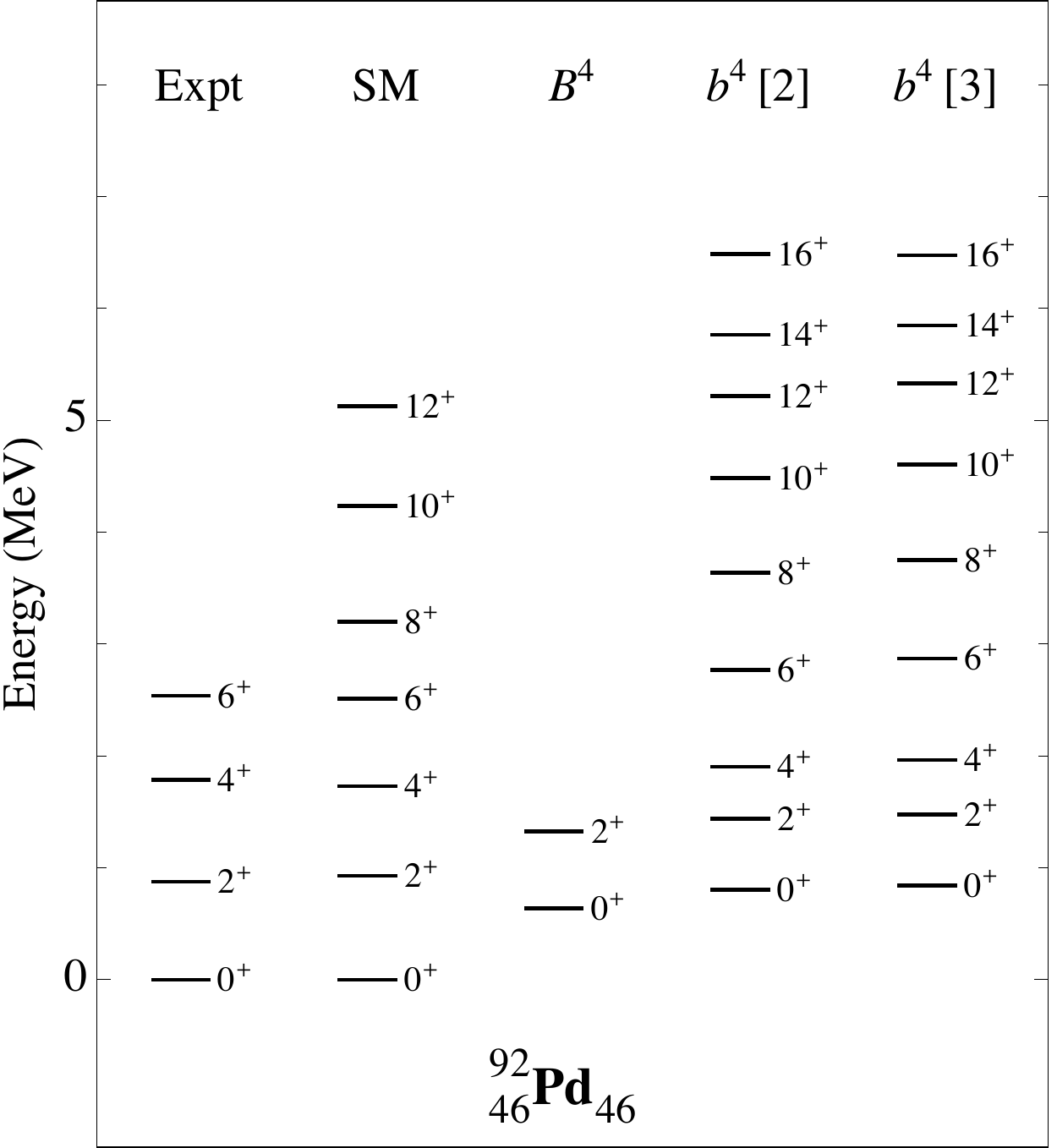}
\caption{
\label{f_spec92}
The yrast spectrum of $^{92}$Pd.
Levels are labelled by their angular momentum and parity $J^\pi$.
The different columns contain the experimental~\cite{Cederwall11} levels (Expt),
the results of the $(1g_{9/2})^8$ shell model (SM) with the SLGT0 interaction,
the lowest eigenvalue of the shell-model hamiltonian in the $B$-pair subspace ($B^4$)
and the lowest eigenvalue of the mapped boson hamiltonian
with up to two-body ($b^4[2]$) and up to three-body ($b^4[3]$) interactions.
The shell-model energy of the $0^+_1$ level is normalized to zero.}
\end{figure}
The energy spectrum of $^{92}$Pd,
calculated in various approximations,
single-$j$ shell model (SM),
shell-model $B$-pair approximation ($B^4$)
and mapped $b$-boson calculation
with up to two-body ($b^4[2]$) and three-body ($b^4[3]$) interactions,
is shown in Fig.~\ref{f_spec92}.
The $B$-pair calculation shows an underbinding of about $0.8$~MeV.
This feature is considerably improved
if $S$ pairs or $s$ bosons are added to the basis~\cite{Zerguine11}.

\begin{figure}
\includegraphics[width=8.5cm]{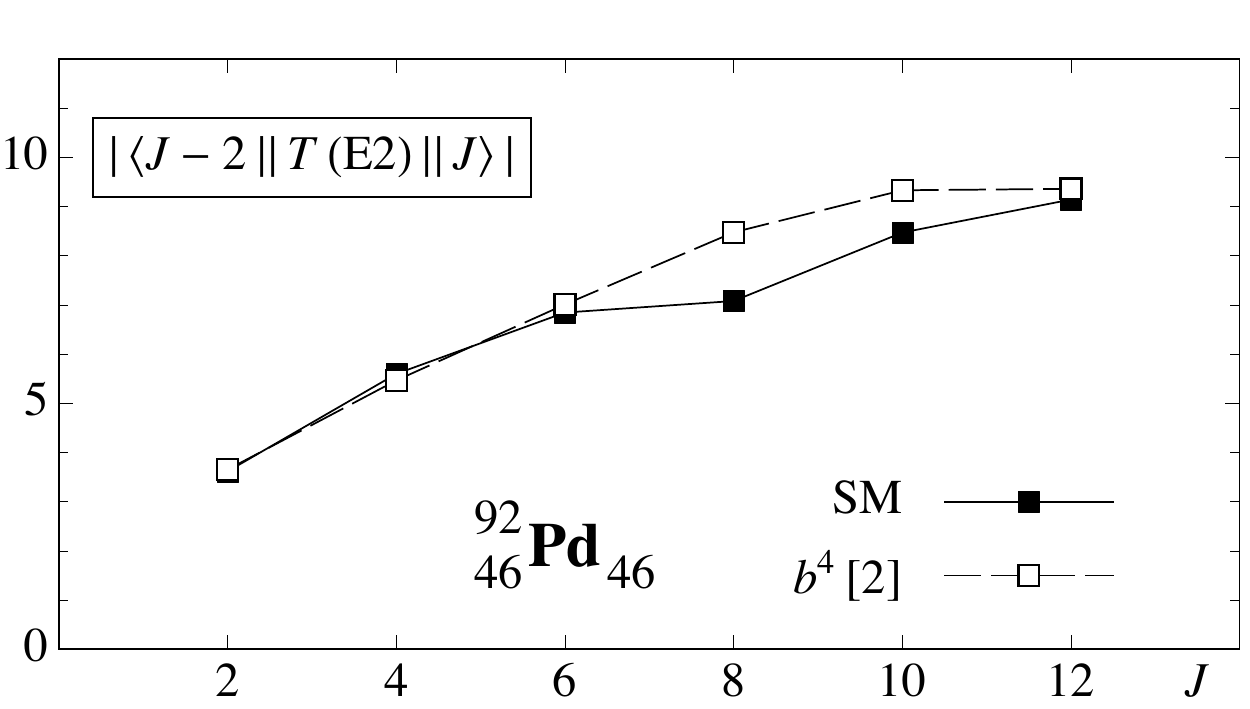}
\caption{
\label{f_e2pd92}
Absolute values of the reduced matrix elements of the E2 operator
for the transitions $J\rightarrow J-2$
between yrast states with $T=0$ ($^{92}$Pd),
calculated in the $(1g_{9/2})^8$ shell model (SM) with the SLGT0 interaction
and compared with the mapped $b$-boson calculation ($b^4[2]$).
Matrix elements are expressed
in units $(e_\nu+e_\pi)l_{\rm ho}^2$.}
\end{figure}
A final illustration of the $b$-boson approximation
is provided in Fig.~\ref{f_e2pd92}
where E2 transition strengths between yrast $(1g_{9/2})^8$ states
as calculated in the shell model
are compared with those obtained with $b$ bosons.
The shell-model reduced matrix elements
are calculated with the operator~(\ref{e_fe2b})
and expressed in units $(e_\nu+e_\pi)l_{\rm ho}^2$.
The reduced matrix elements in the boson approximation
are calculated with the operator~(\ref{e_be2})
with an effective boson charge $e_b$,
determined from the neutron and proton effective charges
according to Eq.~(\ref{e_bfe2}).
A small depletion of the shell-model E2 strength
is perceptible for $J\approx8$ and is absent in the boson calculation.
Apart from this deviation both calculations agree,
indicating once more that the shell-model wave functions
can be adequately represented in terms of a single $b$ boson.
It should be emphasized that,
although the number of $B$-pair states
is but a small subset of the total number of possible shell-model states,
no effective boson charge is needed to arrive
at the agreement found in Fig.~\ref{f_e2pd92}.

\section{Conclusions}
\label{s_conc}
What can be concluded
with regard to the three approximations enounced in Sect.~\ref{s_approx}?
(i) Can the shell-model space be truncated to a single high-$j$ orbit?
(ii) Can the single-$j$ shell space be reduced to one written in terms of aligned $B$ pairs?
(iii) And, finally, can the aligned $B$ pair be replaced by a $b$ boson?
The answer to the question (iii) is unreservedly positive:
owing to its high angular momentum,
the $B$ pair behaves much as a boson.
The mapping from $B$-pair to $b$-boson space
can be made exact by including appropriate interactions between the bosons
but becomes approximate if higher-order interactions are neglected.
The examples of the $1f_{7/2}$ and $1g_{9/2}$ shells
show that two-body interactions between the bosons suffice
and that no higher-order interactions are needed.
The answer to question (ii) is positive for most but not for all yrast states,
that is, most but not all yrast states of $N=Z$ nuclei
can be written in terms of $B$ pairs.
Odd--odd $N=Z$ nuclei in particular behave in a schizophrenic manner
with only a subset of their yrast states
having a sizeable $B$-pair content.
Other states, mostly of low angular momentum,
require the inclusion of low-spin pairs
such as those mapped onto the corresponding bosons of the \mbox{IBM-4}.
In almost all cases, however, a simple interpretation can be given
of yrast states in terms of neutron--proton pairs
and this enables one to intuit the complex spectroscopy of odd--odd $N=Z$ nuclei
and to derive simple parameter-free predictions. 
The validity of the truncation to a single-$j$ shell
depends on specific features of a realistic shell-model hamiltonian
and the answer to question (i) may therefore be different
for the $1f_{7/2}$ and $1g_{9/2}$ shells considered in this review.
A recent large-scale shell-model calculation
with a realistic effective interactions
seems to indicate that the truncation to $1g_{9/2}$
(and therefore the $B$-pair approximation)
is justified in the $A\sim90$--100 region~\cite{Fu13b}.
But, in the end, only the experimental verification of the simple predictions
derived on the basis of the $b$-boson approximation
will be able to tell whether $N=Z$ nuclei exist
with sufficiently isolated single-$j$ shells.

\section*{Acknowledgements}
I wish to thank Salima Zerguine,
my collaborator in the initial stages of this study,
and Gilles de France, Bo Cederwall and Augusto Macchiavelli for many enlightening discussions.

\end{document}